\newcommand{\MMS}{M_{\rm rec}^2}

\newcommand{\BR}{{\cal B}}

\newcommand{\pip}{\pi^+}
\newcommand{\pim}{\pi^-}
\newcommand{\piz}{\pi^0}

\newcommand{\ks}{K_S^0}
\newcommand{\etap}{\eta^{\prime}}
\newcommand{\psp}{\psi(2S)}
\newcommand{\jpsi}{J/\psi}
\newcommand{\EE}{e^+e^-}
\newcommand{\MM}{\mu^+\mu^-}
\newcommand{\LL}{\ell^+\ell^-}
\newcommand{\pp}{\pi^+\pi^-}

\newcommand{\kk}{K^+K^-}

\newcommand{\ra}{\rightarrow}
\newcommand{\beq}{\begin{equation}}
\newcommand{\eeq}{\end{equation}}
\newcommand{\beqy}{\begin{eqnarray}}
\newcommand{\eeqy}{\end{eqnarray}}
\newcommand{\bitm}{\begin{itemize}}
\newcommand{\eitm}{\end{itemize}}

\newcommand{\GG}{\gamma\gamma}
\newcommand{\ppp}{\pi^+\pi^-\pi^0}

\newcommand{\gevcs}{{\rm GeV}/c^2}
\newcommand{\mev}{\rm MeV}
\newcommand{\mevcs}{{\rm MeV}/c^2}

\newcommand{\ev}{\rm eV}

\newcommand{\op}{\omega\phi}
\newcommand{\oo}{\omega\omega}
\newcommand{\sumpt}{|\sum{\vec{p}_{t}^{\,*}}|}
\newcommand{\kkp}{K\bar{K}\pi}
\newcommand{\kskp}{K_SK^+\pi^-}
\newcommand{\kkpiz}{\kk\piz}
\newcommand{\kskppp}{K_SK^+\pp\pi^-}
\newcommand{\kkpppiz}{\kk\pp\piz}

\newcommand{\pppppp}{3(\pp)}

\newcommand{\etapp}{\eta\pp}

\newcommand{\etac}{\eta_c}
\newcommand{\mevcc}{\mathrm{MeV}/c^2}

\newcommand{\etacp}{\eta_{c}(2S)}

\documentclass{ws-ijmpcs}

\begin{document}

\markboth{Chengping Shen}
{Charmonium and light hadron spectroscopy}

%
\catchline{}{}{}{}{}
%

\title{Charmonium and light hadron spectroscopy}

\author{Chengping Shen}

\address{
School of Physics and Nuclear Energy Engineering, Beihang University\\
Beijing, 100191, China\\
shencp@ihep.ac.cn}

\maketitle


\begin{abstract}
In this report I review some results on the charmonium and light hadron spectroscopy
mainly from BESIII and Belle experiments.
For the charmonium, the contents include the observation of $\psi(4040)/\psi(4160) \to \eta \jpsi$,
the measurements of the $\eta_c/\eta_c(2S)$ resonance parameters and their decays,
the evidence of the $\psi_2(1^3D_2)$ state in the $\chi_{c1}\gamma$ mass spectrum.
For the light hadron spectroscopy, the  contents include the $X(1835)$ research in
$e^+e^- \to \jpsi + X(1835)$ and $\gamma \gamma \to \eta' \pp$ processes, and the analysis
of the $\eta \eta$, $\omega \phi$, $\phi\phi$ and $\omega \omega$ mass spectra in low mass region.

\end{abstract}

\ccode{PACS numbers: 14.40.Pq, 13.25.-k, 13.25.Gv}


\section{$\psi(4040)/\psi(4160) \to \eta \jpsi$ }	

Experimentally well established structures
$\psi(4040)$, $\psi(4160)$, and
$\psi(4415)$ resonances above
the $D\bar{D}$ production threshold are of great interest but not well
understood, even decades after their first
observation.

BESIII accumulated a $478$~pb$^{-1}$ data sample at a center-of-mass (CMS)
energy of $\sqrt{s}=4.009$~GeV. Using this data sample, the processes
$\EE\to\eta\jpsi$ and $\pi^0\jpsi$ cross section are measured~\cite{liu}.
In this analysis, the $\jpsi$ is reconstructed through its
decays into lepton pairs while $\eta/\pi^0$ is
reconstructed in the $\GG$ final state. After imposing all of some selection criteria,
a clear $\jpsi$ signal is observed in the $\MM$ mode while
indications of a peak
around 3.1~GeV/c$^2$ also exist in the $\EE$ mode.

\begin{figure*}[htbp]
\begin{center}
\includegraphics[height=2.95cm]{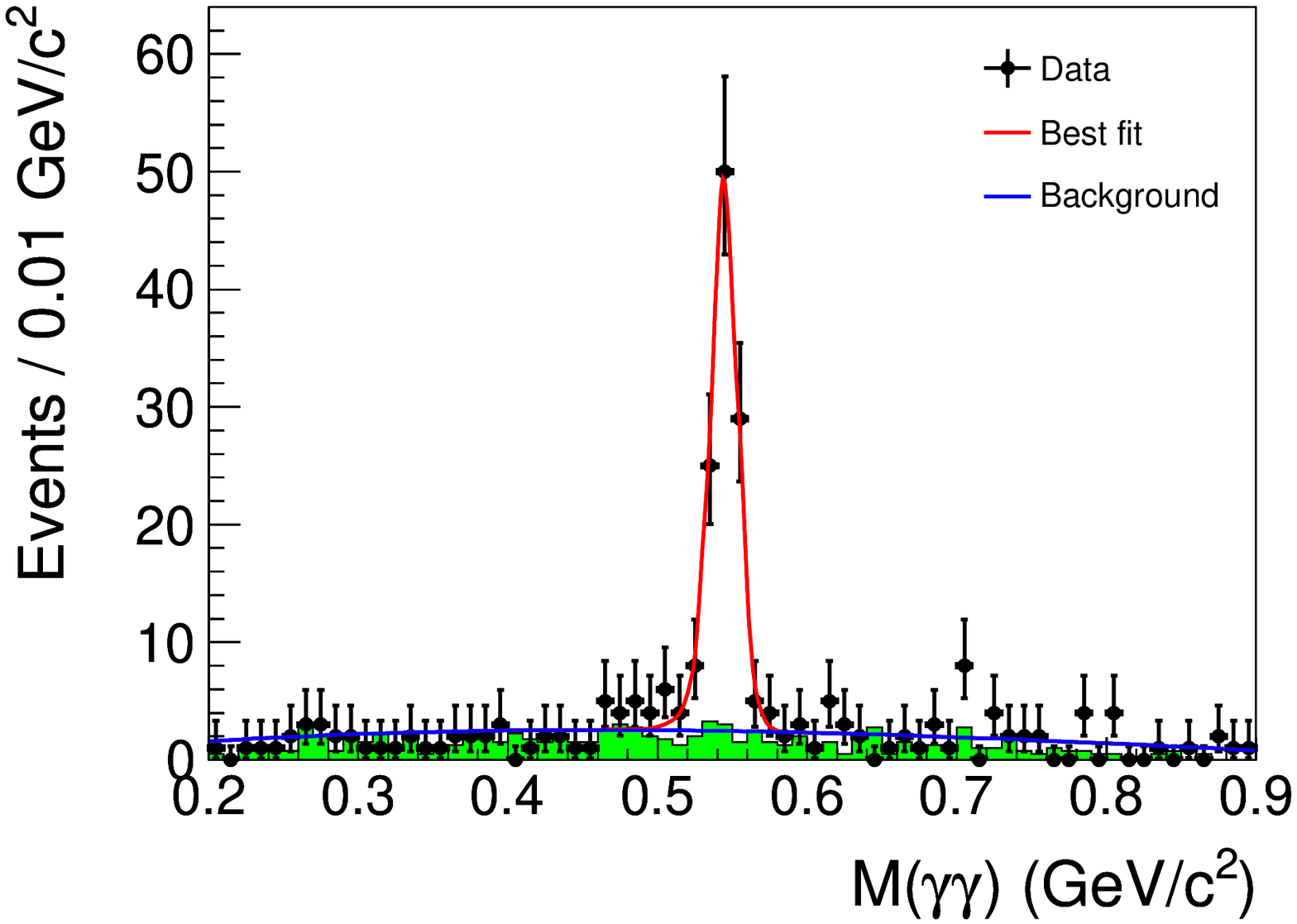}
\includegraphics[height=2.95cm]{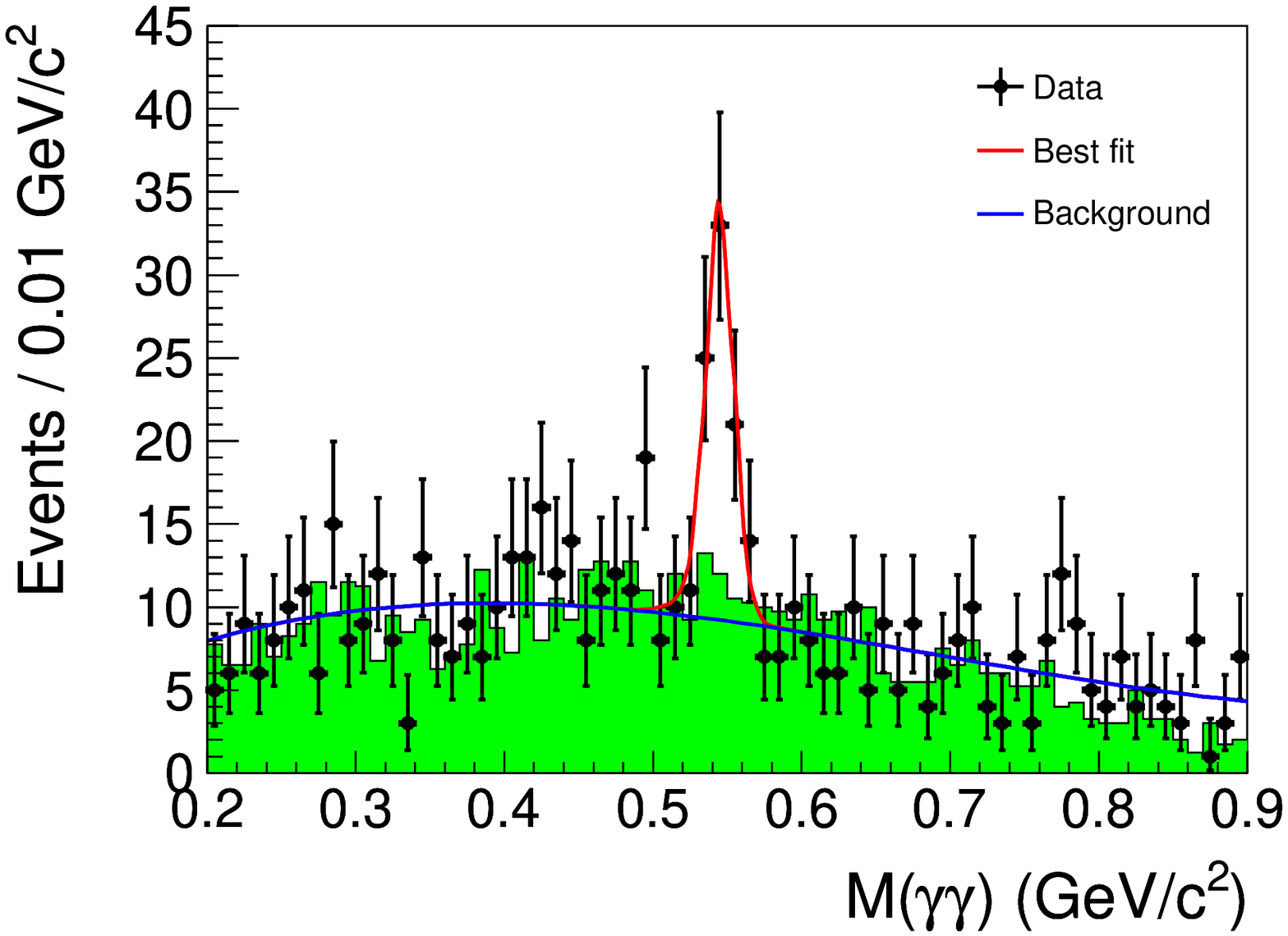}
\includegraphics[height=2.95cm]{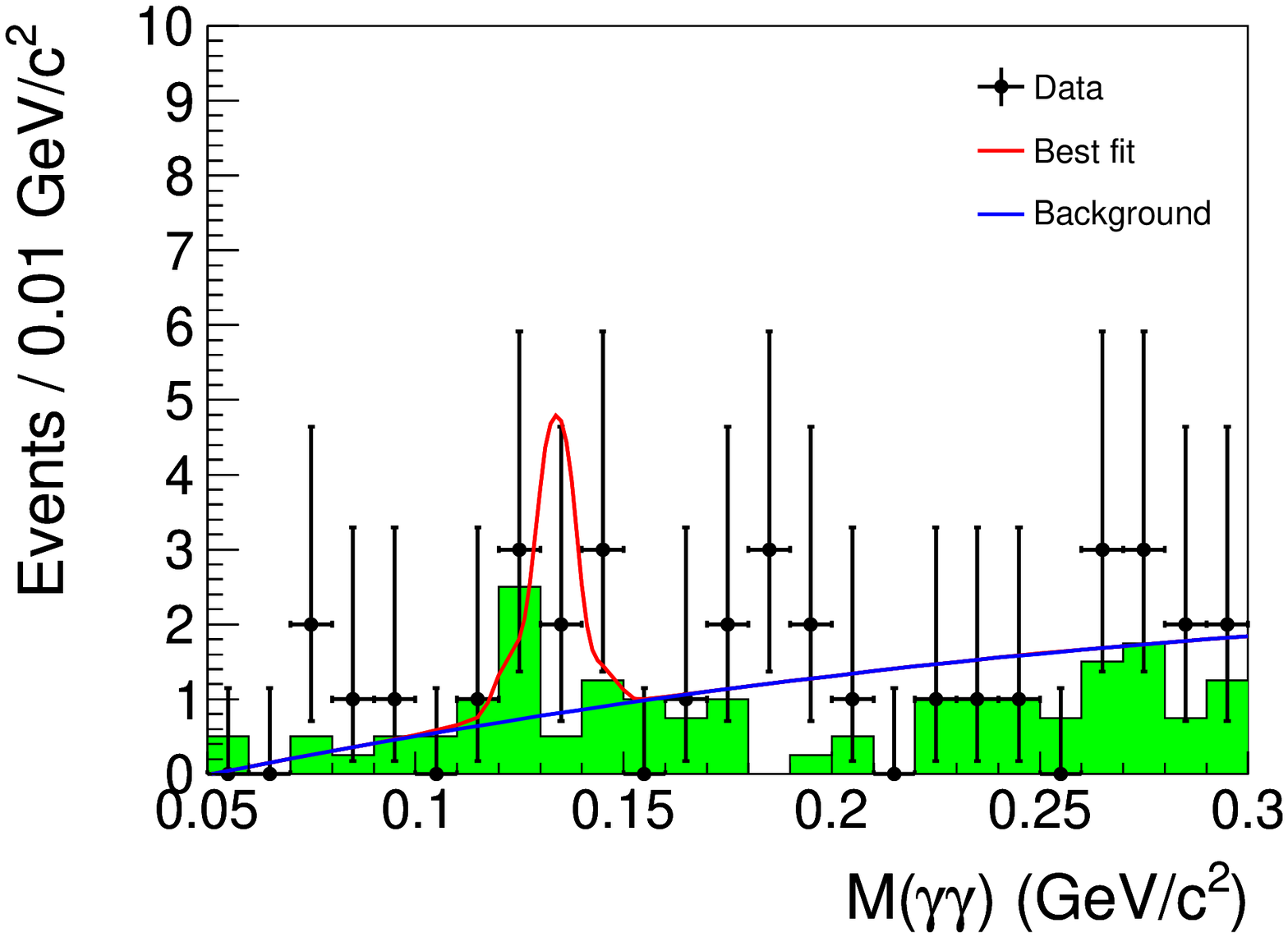}
\caption{ Distributions of $M(\gamma \gamma)$ between 0.2~GeV/c$^2$ and 0.9~GeV/c$^2$
for $\jpsi\to \MM$ (left panel) and for
$\jpsi\to \EE$ (middle panel) and distribution of $M(\gamma \gamma)$ below 0.3~GeV/c$^2$
for $\jpsi\to \MM$ (right panel). Dots with error
bars are data in $\jpsi$ mass signal region, and the green shaded
histograms are from normalized $\jpsi$ mass sidebands. The curves
show the total fit and the background term.} \label{fit-mgg}
\end{center}
\end{figure*}

A significant $\eta$ signal is observed
in $M(\GG)$ in both $\jpsi\to \MM$ and
$\jpsi\to \EE$, as shown in Fig.~\ref{fit-mgg}. No
significant $\pi^0$ signal is observed. The $M(\GG)$ invariant mass distributions are fitted using an unbinned
maximum likelihood method.  For the $\eta$
signal, the statistical significance is larger than $10\sigma$ while
that for the $\pi^0$ signal is only $1.1\sigma$.
The Born cross section for $\EE\to \eta\jpsi$ is measured to be
$(32.1\pm 2.8 \pm 1.3)$~pb, and the Born cross section is found to be less
than 1.6~pb at the 90\% confidence level (C.L.) for $\EE\to
\pi^0\jpsi$.

Belle used 980~fb$^{-1}$ data to study the process
$\EE \to \eta\jpsi$ via ISR~\cite{wangxl}. $\eta$ is reconstructed in the $\gamma \gamma$
and $\pi^+ \pi^- \pi^0$ final states. Due to the high background level from Bhabha
scattering, the $\jpsi\to \EE$ mode is not used in conjunction
with the decay mode $\eta\to \gamma \gamma$.

\begin{figure}[htbp]
\begin{center}
\includegraphics[width=5.0cm, angle=-90]{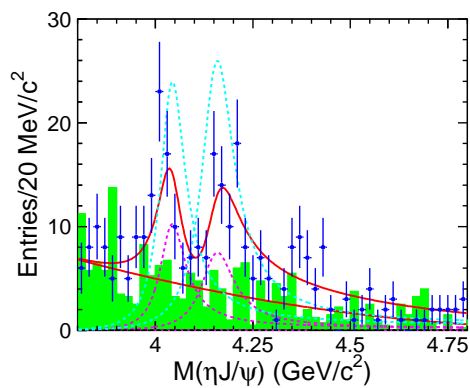}
\caption{The $\eta\jpsi$ invariant mass distribution and the fit
results. The points with error bars show the data while the shaded
histogram is the normalized $\eta$ and $\jpsi$ background from the
sidebands. The curves show the best fit on signal candidate events
and sideband events simultaneously and the contribution from each
Breit-Wigner component.The dashed curves at each peak show the
two solutions.} \label{fit}
\end{center}
\end{figure}

Clear $\eta$ and $\jpsi$ signals could be observed.
A dilepton pair is considered as a $\jpsi$ candidate
if $M_{\LL}$ is within $\pm 45~\mevcs$ of the $\jpsi$ nominal mass.
The $\eta$ signal region is defined as $M_{\ppp} \in [0.5343,
0.5613]~\gevcs$ and $M_{\GG}\in [0.5,0.6]~\gevcs$.
$-1~(\gevcs)^2 < \MMS < 2.0~(\gevcs)^2$ is required to select ISR candidates, where $\MMS$ is
the square of the mass recoiling against the $\eta\jpsi$ system.
After event selections, an unbinned maximum likelihood fit is performed to the mass
spectra $M_{\eta\jpsi}\in [3.8,4.8]~\gevcs$ from the  signal
candidate events and $\eta$ and $\jpsi$ sideband events
simultaneously, as shown in Fig.~\ref{fit}. The fit to the signal
events includes two coherent $P$-wave Breit-Wigner functions,
$BW_1$ for $\psi(4040)$ and $BW_2$ for $\psi(4160)$, and an
incoherent second-order polynomial background.
Statistical significance is $6.5\sigma$ for $\psi(4040)$ and $7.6\sigma$ for
$\psi(4160)$. There are two solutions with equally good fit quality:
$\BR(\psi(4040)\to\eta\jpsi)\cdot\Gamma_{\EE}^{\psi(4040)} =
(4.8\pm0.9\pm1.4)~\ev$ and
$\BR(\psi(4160)\to\eta\jpsi)\cdot\Gamma_{\EE}^{\psi(4160)} =
(4.0\pm0.8\pm1.4)~\ev$ for one solution and
$\BR(\psi(4040)\to\eta\jpsi)\cdot\Gamma_{\EE}^{\psi(4040)} =
(11.2\pm1.3\pm1.9)~\ev$ and
$\BR(\psi(4160)\to\eta\jpsi)\cdot\Gamma_{\EE}^{\psi(4160)} =
(13.8\pm1.3\pm2.0)~\ev$ for the other solution, where the first
errors are statistical and the second are systematic. The partial widths to $\eta\jpsi$ are found to be
about $1~\mev$.

\section{Some results on $\eta_c$ and $\eta_c(2S)$}

The $\etac$ mass and width have large uncertainties.
The measured results of the $\etac$ mass and width from $\jpsi$ radiative transitions
and two-photon fusion and $B$ decays have large inconsistence.
The most recent study by the CLEO-c experiment, using both $\psp \to
\gamma\etac$ and $\jpsi\to \gamma\etac$, pointed out a
distortion of the $\etac$ line shape in $\psp$ decays.

With a $\psp$ data sample of $1.06\times 10^8$ events, BESIII
reported measurements of the $\etac$ mass and
width using the radiative transition $\psp \to \gamma \etac$~\cite{bes3-etac}.
Six modes are used to
reconstruct the $\etac$: $\kskp$, $\kkpiz$, $\etapp$,  $\kskppp$,
$\kkpppiz$, and $\pppppp$, where the $\ks$ is reconstructed in
$\pp$, and the $\eta$ and $\piz$ in $\gamma\gamma$ decays.

\begin{figure*}[htbp]
\begin{center}
\includegraphics[width=4.1cm]{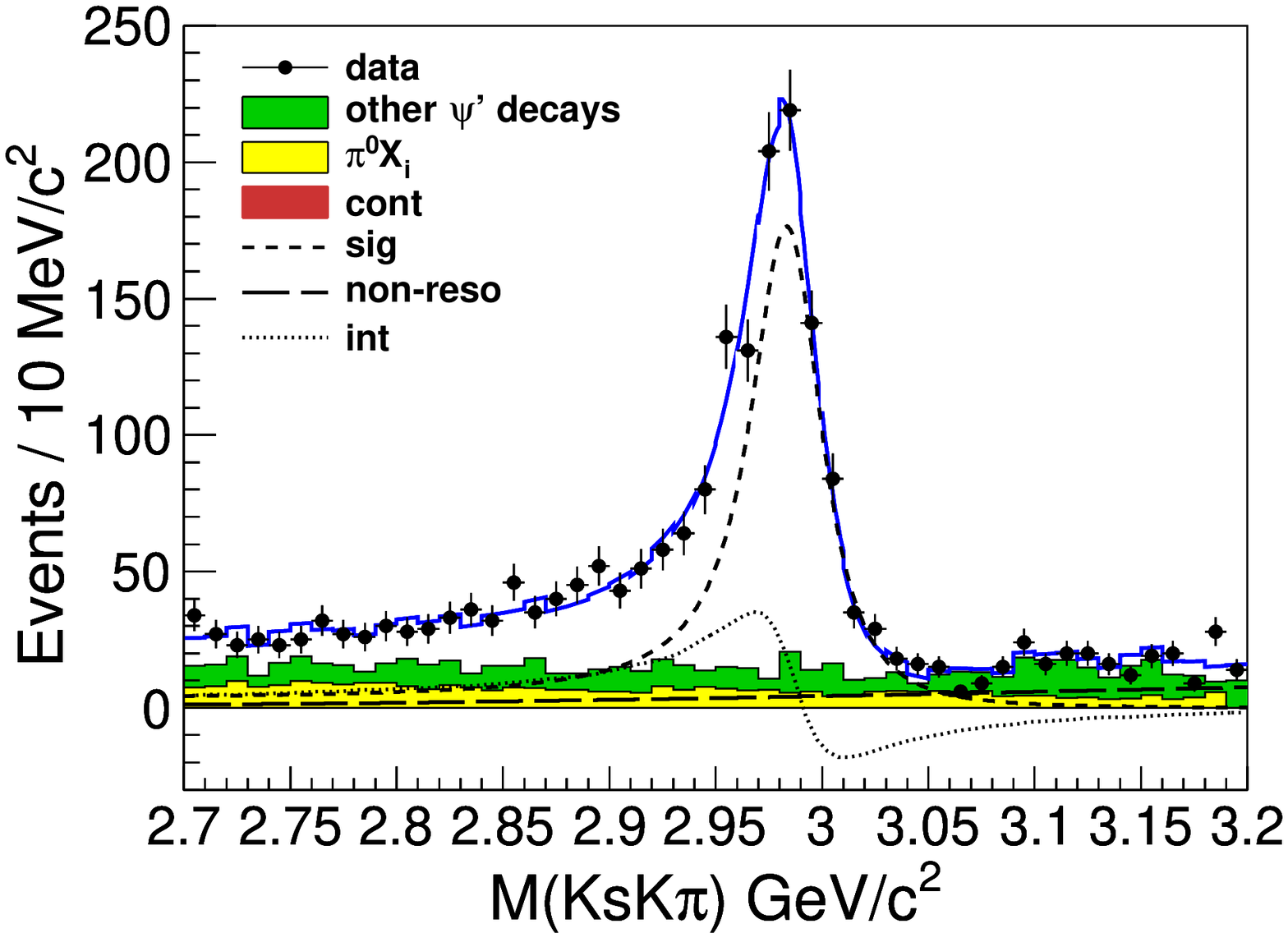}
\includegraphics[width=4.1cm]{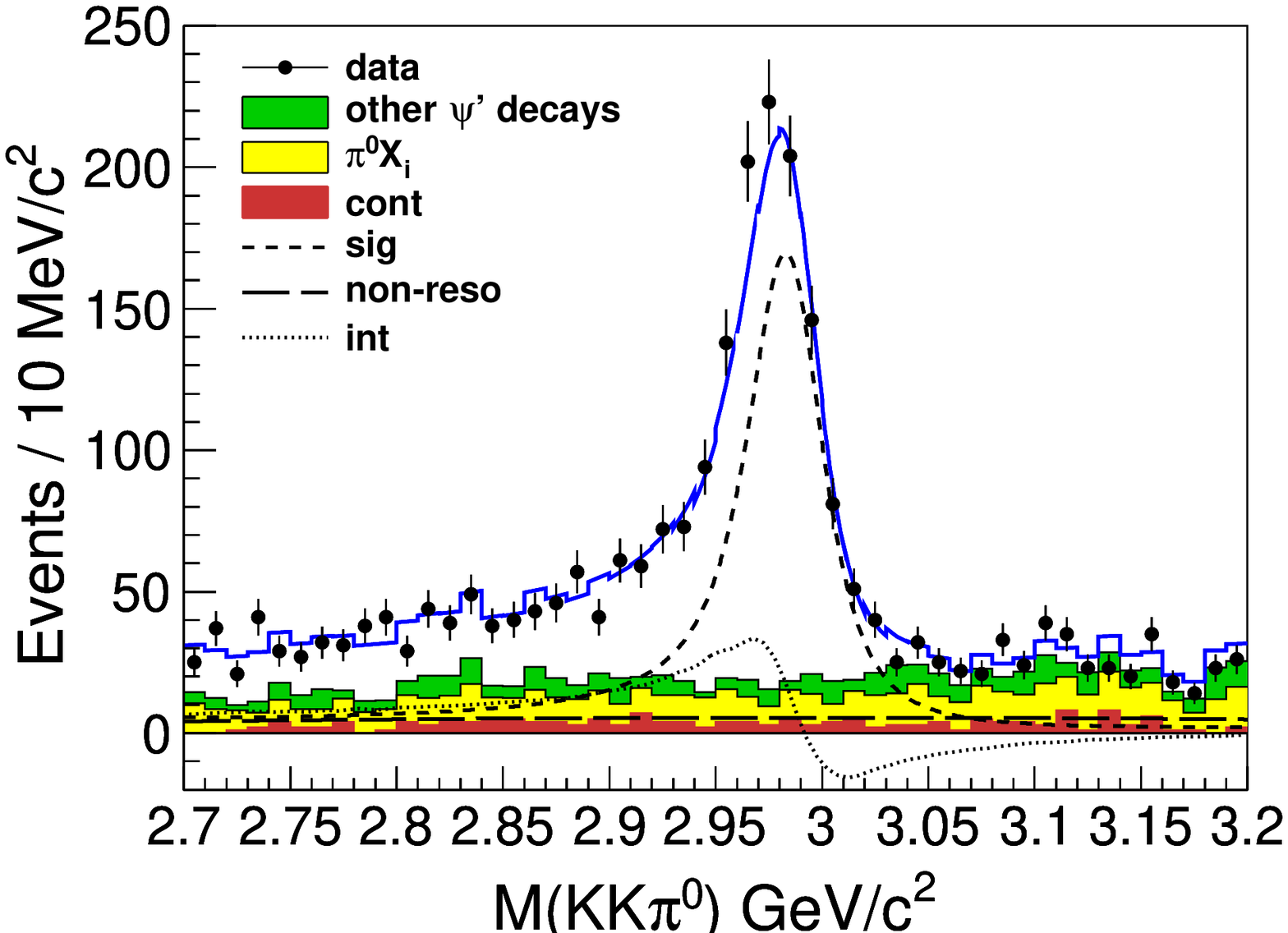}
\includegraphics[width=4.1cm]{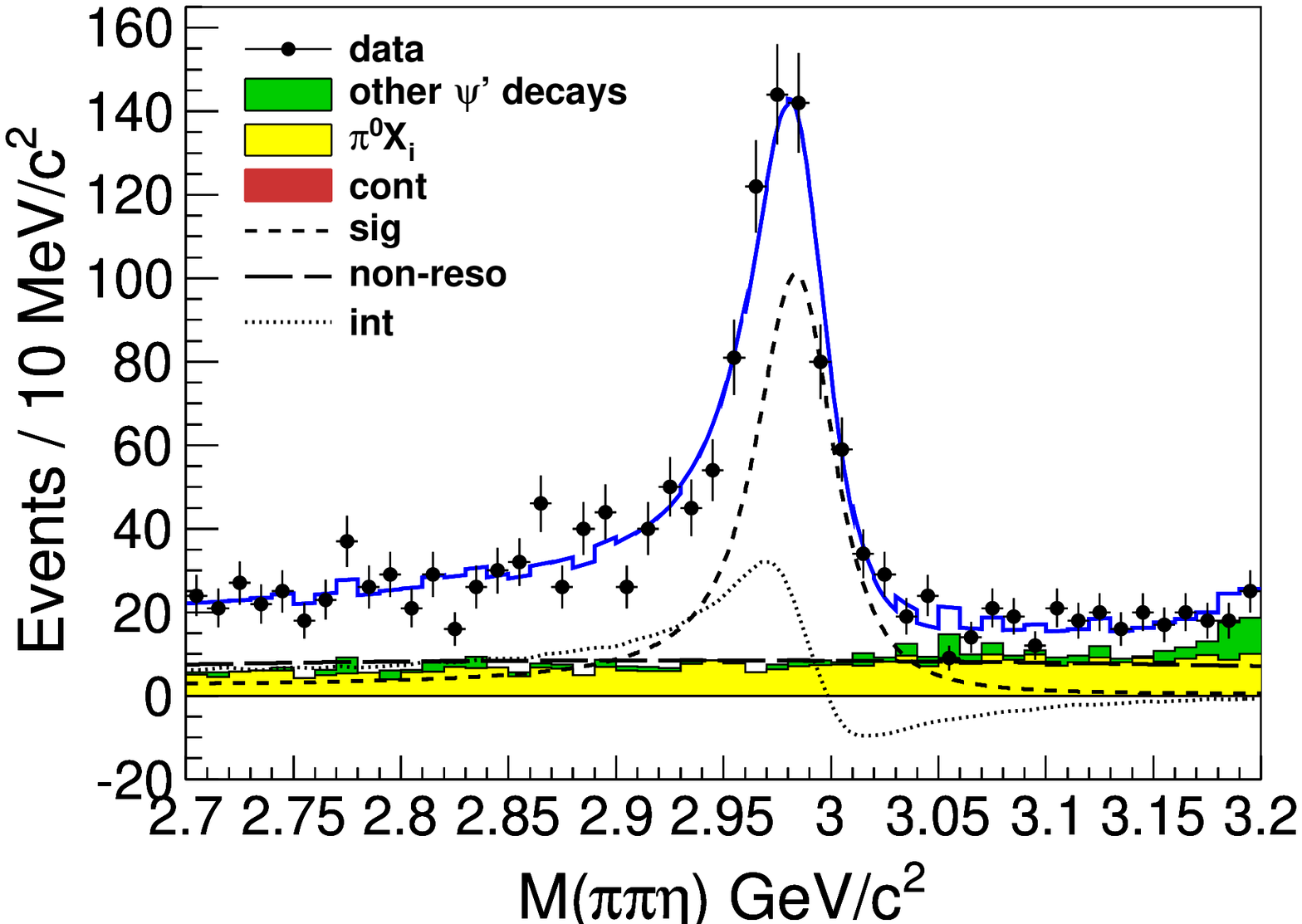}
\includegraphics[width=4.1cm]{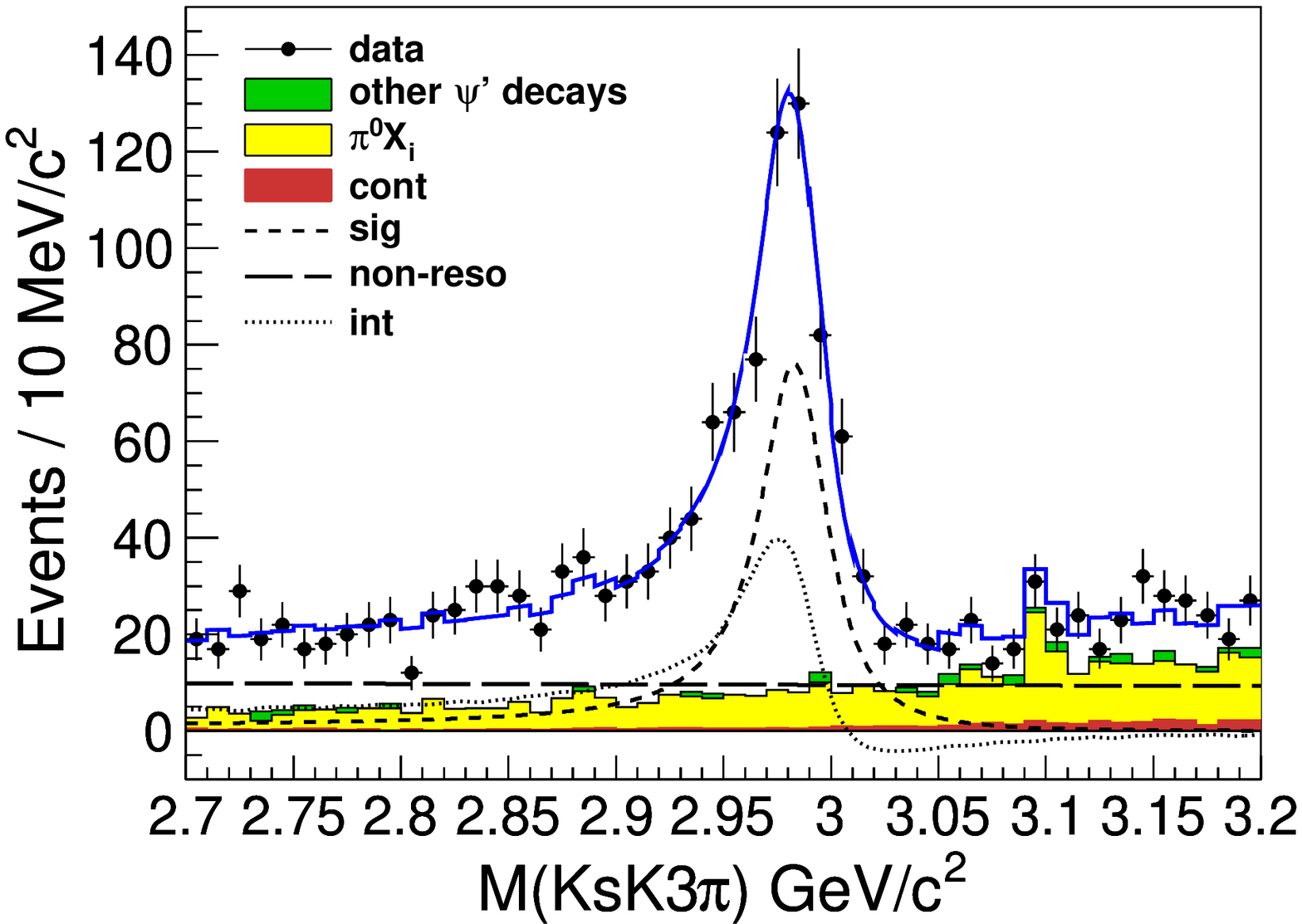}  \hspace{0.07cm}
\includegraphics[width=4.1cm]{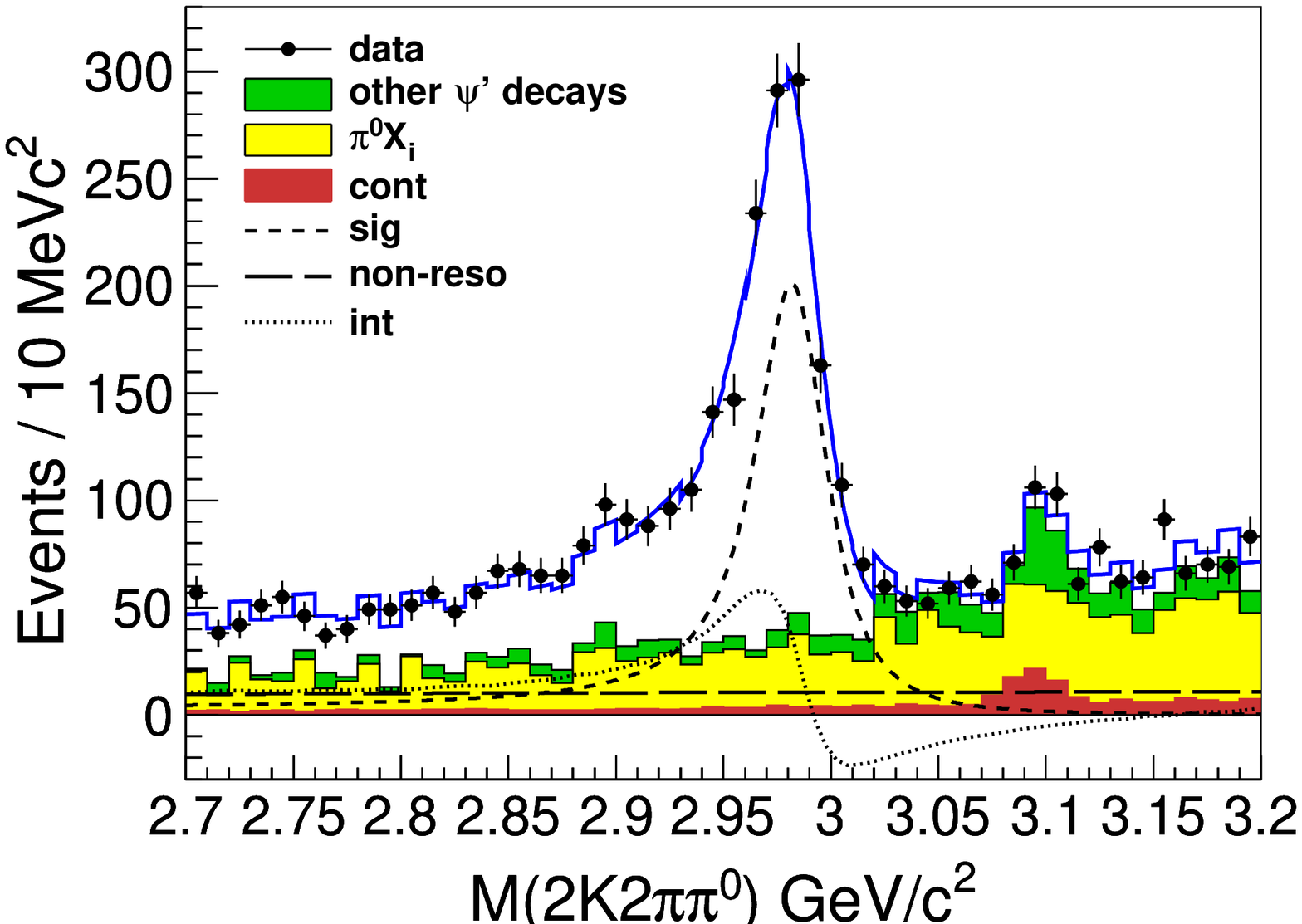}
\includegraphics[width=4.1cm]{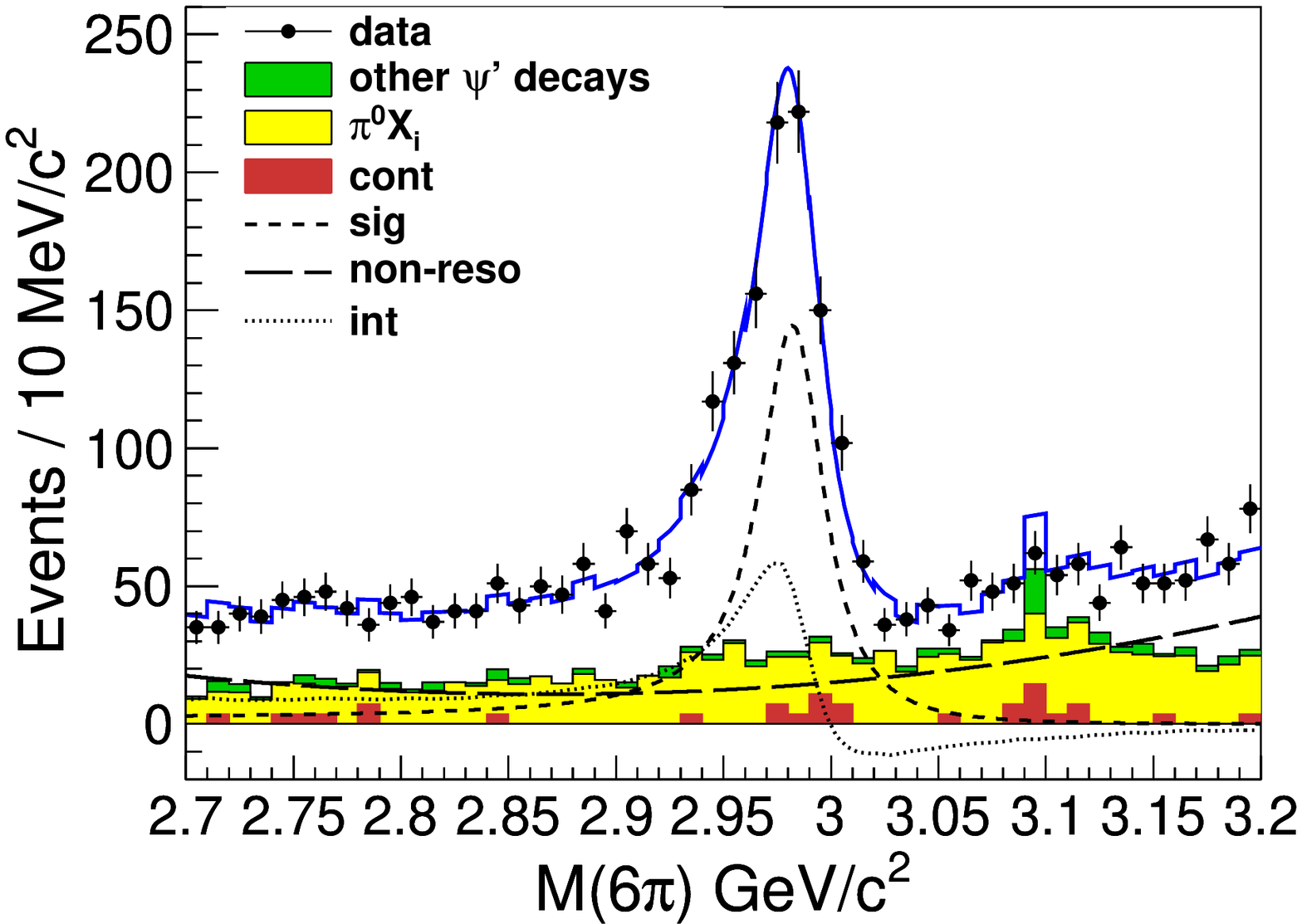}
\end{center}
\caption{The $M(X_i)$ invariant mass distributions for the decays
  $\kskp$, $\kkpiz$, $\etapp$, $\kskppp$, $\kkpppiz$ and $\pppppp$,
    respectively, with the fit results (for the constructive solution)
  superimposed. Points are data and the various curves are the total fit
  results. Signals are shown as short-dashed lines; the non-resonant
  components as long-dashed lines; and the interference between them
  as dotted lines.
  Shaded histograms are (in red/yellow/green) for (continuum/$\pi^0
  X_i$/other $\psp$ decays) backgrounds.
  The continuum backgrounds for  $\kskp$ and $\etapp$ decays are
  negligible.}
  \label{fig:metac}
\end{figure*}

Figure~\ref{fig:metac} shows the $\etac$ invariant mass
distributions for selected $\etac$ candidates, together with the
estimated backgrounds. A clear $\etac$
signal is evident in every decay mode.
Assuming 100\% interference between the
$\etac$ and the non-resonant amplitude, an unbinned
simultaneous maximum likelihood fit was performed.
In the fit, the $\etac$ mass, width, and relative phases
are free parameters, and the mass and width are constrained to be the
same for all decay modes. Two solutions of relative phase are found for every decay mode,
one represents constructive interference, the other for
destructive. The measured mass is $M = 2984.3\pm 0.6 (stat.)\pm 0.6(syst.)~\mevcc$ and width
$\Gamma = 32.0  \pm 1.2 (stat.)\pm 1.0(syst.)~\mev$. The interference is significant,
which indicates previous measurements of the $\etac$ mass and width via radiative
transitions may need to be rechecked. The results are consistent with
that from photon-photon fusion and $B$ decays; this may partly clarify the discrepancy puzzle.

Similarly the properties of the $\etacp$ are not well-established either.
The $\etacp$ was first observed by the Belle collaboration in the
process $B^\pm\to K^\pm \etacp$, $\etacp\to K_S^0K^\pm
\pi^\mp$.  It was confirmed in the
two-photon production of $K_S^0K^\pm
\pi^\mp$, and in the double-charmonium production process
$e^+e^-\to J/\psi c\bar{c}$. Combining the world-average
values with the most recent results from Belle and
BaBar on two-photon fusion into hadronic final states other than
$K_S^0K^\pm
\pi^\mp$, one
obtains updated averages of the $\etacp$ mass and width of
$3637.7\pm 1.3~{\rm MeV}/c^2$ and $10.4\pm 4.2~{\rm MeV}$,
respectively. $\etacp$ was also observed in six-prong final states in two-proton
processes including $3(\pi^+\pi^-)$, $K^+K^-2(\pp)$, $2(K^+K^-)\pp$,
$K^{0}_{S} K^{\pm}\pi^{\mp}\pi^{+}\pi^{-}$ by Belle collaboration.
The measured averaged mass and width of $\etacp$ are
$3636.9\pm1.1\pm2.5\pm5.0$ MeV/$c^2$ and $9.9\pm3.2\pm2.6\pm2.0$ MeV/$c^2$.
The results were reported in ICHEP2010 meeting, but the results are still
preliminary up to date.

Recently BESIII collaboration searched for the M1 radiative transition
$\psp \ra \gamma \etacp$ by reconstructing the exclusive
$\etacp \ra K^{0}_{S} K^{\pm}\pi^{\mp}\pi^{+}\pi^{-}$ decay using
1.06 $\times$ $10^{8}$ $\psp$ events~\cite{bes3-etac2s}.

\begin{figure*}[htbp]
\begin{center}
\includegraphics[width=3.5in,height=2.5in]{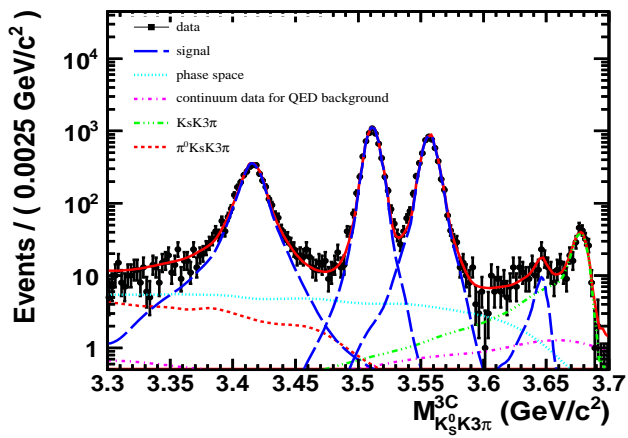}
\end{center}
\caption{
The results of fitting the mass spectrum for $\chi_{cJ}$ and $\etacp$. The black dots are the data,
the blue long-dashed line shows the $\chi_{cJ}$ and $\etacp$ signal shapes, the cyan dotted line represents
the phase space contribution, the violet dash-dotted line shows the continuum data contribution, the green
dash-double-dotted line shows the contribution of $\psp \to K^{0}_{S} K^{\pm}\pi^{\mp}\pi^{+}\pi^{-}$, and the red dashed line is the contribution
of $\psp \to \pi^{0} K^{0}_{S} K^{\pm}\pi^{\mp}\pi^{+}\pi^{-}$.
}
\label{fig:fitting_total}
\end{figure*}

The final mass spectrum of $K^{0}_{S} K^{\pm}\pi^{\mp}\pi^{+}\pi^{-}$
and the fitting results are shown in Fig.~\ref{fig:fitting_total}.
The fitting function consists of the following components:
$\etacp$, $\chi_{cJ}(J= 0, 1, {\rm and}~2)$ signals and
$\psp \ra K^{0}_{S} K^{\pm}\pi^{\mp}\pi^{+}\pi^{-}$,
$\psp \ra \pi^{0} K^{0}_{S} K^{\pm}\pi^{\mp}\pi^{+}\pi^{-}$, ISR,
and phase space backgrounds.
The result for the yield of $\etacp$ events is $57\pm17$ with a significance of 4.2$\sigma$.
The measured mass of the $\etacp$ is 3646.9 $\pm 1.6(stat.) \pm 3.6(syst.)$ $\mevcc$, and the width
is 9.9 $\pm 4.8(stat.) \pm 2.9(syst.)$ $\mevcc$. The product branching fraction is measured to be
$\BR(\psp \ra \gamma \etacp) \times \BR(\etacp \ra K^{0}_{S} K^{\pm}\pi^{\mp}\pi^{+}\pi^{-})$ =
(7.03 $\pm 2.10(stat.) \pm 0.70(syst.)$) $\times$ $10^{-6}$.
This measurement complements a previous BESIII measurement of $\psp \ra \gamma\etacp$ with $\etacp \ra \kskp$ and $\kkp$.

\section{Evidence  of the $1^3D_2~c\bar{c}$ state (X(3823))}

During the last decade, a number of new charmonium
($c\bar{c}$)-like states were observed, many of which are candidates for
exotic states. The observation of a
$D$-wave $c\bar{c}$ meson and its decay modes would test phenomenological
models.  The undiscovered $1^3D_2~c\bar{c}$ $(\psi_2$) and $1^3D_3~c\bar{c}~(\psi_3)$ states are
expected to have significant branching fractions to $\chi_{c1}\gamma$ and
$\chi_{c2}\gamma$, respectively. So Belle used $772\times 10^{6}$ $B\overline{B}$  events
to search for the possible structures in $\chi_{c1} \gamma$ and $\chi_{c2} \gamma$ mass spectra
in the processes $B \to \chi_{c1} \gamma K$ and
$B \to \chi_{c2} \gamma K$ decays, where the $\chi_{c1}$ and $\chi_{c2}$
decay to $J/\psi \gamma$~\cite{x3823}. The $J/\psi$ meson is reconstructed via its decays to $\ell^+\ell^-$
($\ell =$ $e$ or $\mu$).

\begin{figure}[h!]
\begin{center}
\includegraphics[height=55mm,width=80mm]{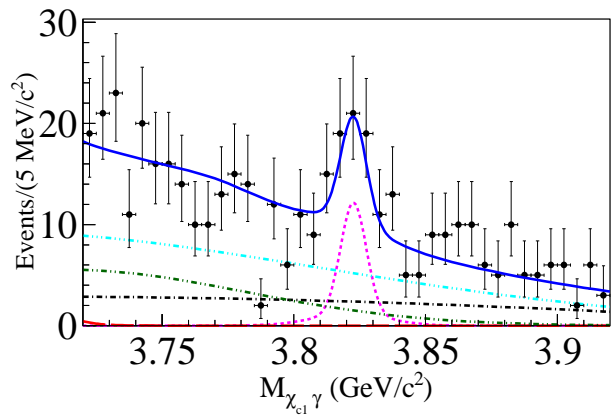}
\caption{\label{fig:sim} Two-dimensional unbinned extended
maximum likelihood fit projection of $M_{\chi_{c1}\gamma}$
distribution for the simultaneous fit of  $B^{\pm} \to (\chi_{c1} \gamma) K^{\pm}$
and  $B^{0} \to (\chi_{c1} \gamma) K_S^{0}$  decays
for $M_{\rm bc} > 5.27 $ GeV$/c^2$. }
\end{center}
\end{figure}

The  $M_{\chi_{c1}\gamma}$ distribution from $B^{\pm} \to (\chi_{c1} \gamma) K^{\pm}$
and  $B^{0} \to (\chi_{c1} \gamma) K_S^{0}$  decays was shown in Fig.~\ref{fig:sim},
where there is a significant narrow peak at
3823 MeV/$c^2$, denoted   hereinafter as $X(3823)$.  No signal
of $X(3872) \to \chi_{c1}\gamma$ is seen.
To extract the mass of the $X(3823)$,
a simultaneous fit to $B^{\pm} \to (\chi_{c1}\gamma) K^{\pm}$
and $B^0 \to (\chi_{c1}\gamma) K_S^0$ is performed, assuming that
$\mathcal{B}(B^\pm \to X(3823) K^\pm)/\mathcal{B}(B^0 \to X(3823)K^0)$ =
$\mathcal{B}(B^\pm \to \psi' K^\pm)/\mathcal{B}(B^0 \to \psi' K^0)$.
The mass of the $X(3823)$ is measured to be
$3823.1\pm1.8({stat.})\pm 0.7{(syst.)}$ MeV$/c^2$ and
signal significance is estimated to be 3.8$\sigma$ with systematic  uncertainties included.
The measured branching fraction product
$\mathcal{B}(B^{\pm} \to X(3823) K^{\pm})  \mathcal{B}(X(3823) \to \chi_{c1}\gamma)$
is  $(9.7 \pm 2.8 \pm 1.1)\times 10^{-6}$. No
evidence is found for $X(3823)\to \chi_{c2}\gamma$.
 The properties of the $X(3823)$ are consistent with  those
expected for the $\psi_2~(1 ^3D_2~ c\bar{c})$ state.

\section{Search for $X(1835)$}

In the radiative decay $J/\psi\to\gamma \pi^+\pi^-\eta'$, the BESII
Collaboration observed a resonance, the $X(1835)$,
with a statistical significance of 7.7$\sigma$.
Recently the structure has been confirmed by BESIII
in the same process with
$2.25\times 10^8$ $J/\psi$ events.

Many theoretical models have been proposed to interpret its underlying structure.
Some interpret $X(1835)$ as radial excitation of $\eta^{'}$,
a $p\bar{p}$ bound state, a glueball
candidate, or a $\eta_{c}$-glueball mixture.

Belle first tried to search for the $X(1835)$
in the two-photon process  $\gamma \gamma \to \etap\pip\pim$
using a 673 fb$^{-1}$ data sample with $\etap\to\eta\pip\pim$,
and $\eta\to \gamma\gamma$~\cite{zhangcc}.

Significant background reduction is achieved
by applying a $\sumpt$  requirement ($|\sum{\vec{p}_{t}^{\,*}}| < 0.09~$ GeV$/c$), which is determined by taking the absolute value of the vector sum of the
transverse momenta of $\etap$ and the $\pip\pim$ tracks
in the $e^+e^-$ center-of-mass system.
The $\sumpt$ distribution for the signal peaks
at small values, while that for both backgrounds
decreases toward $|\sum{\vec{p}_{t}^{\,*}}| = 0$
due to vanishing phase space.

\begin{figure}
\begin{center}
\includegraphics[width=6cm]{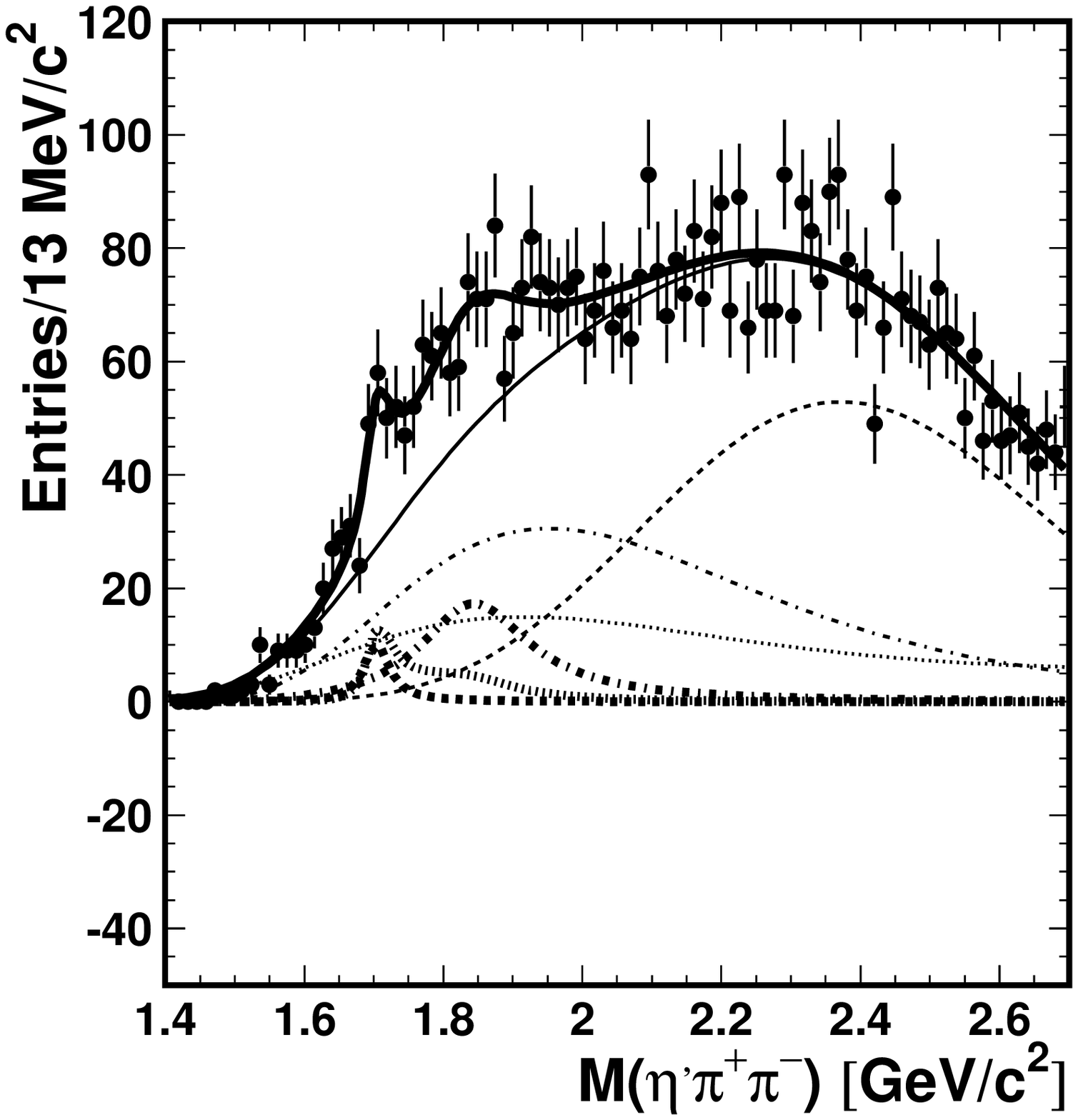}
\includegraphics[width=6cm]{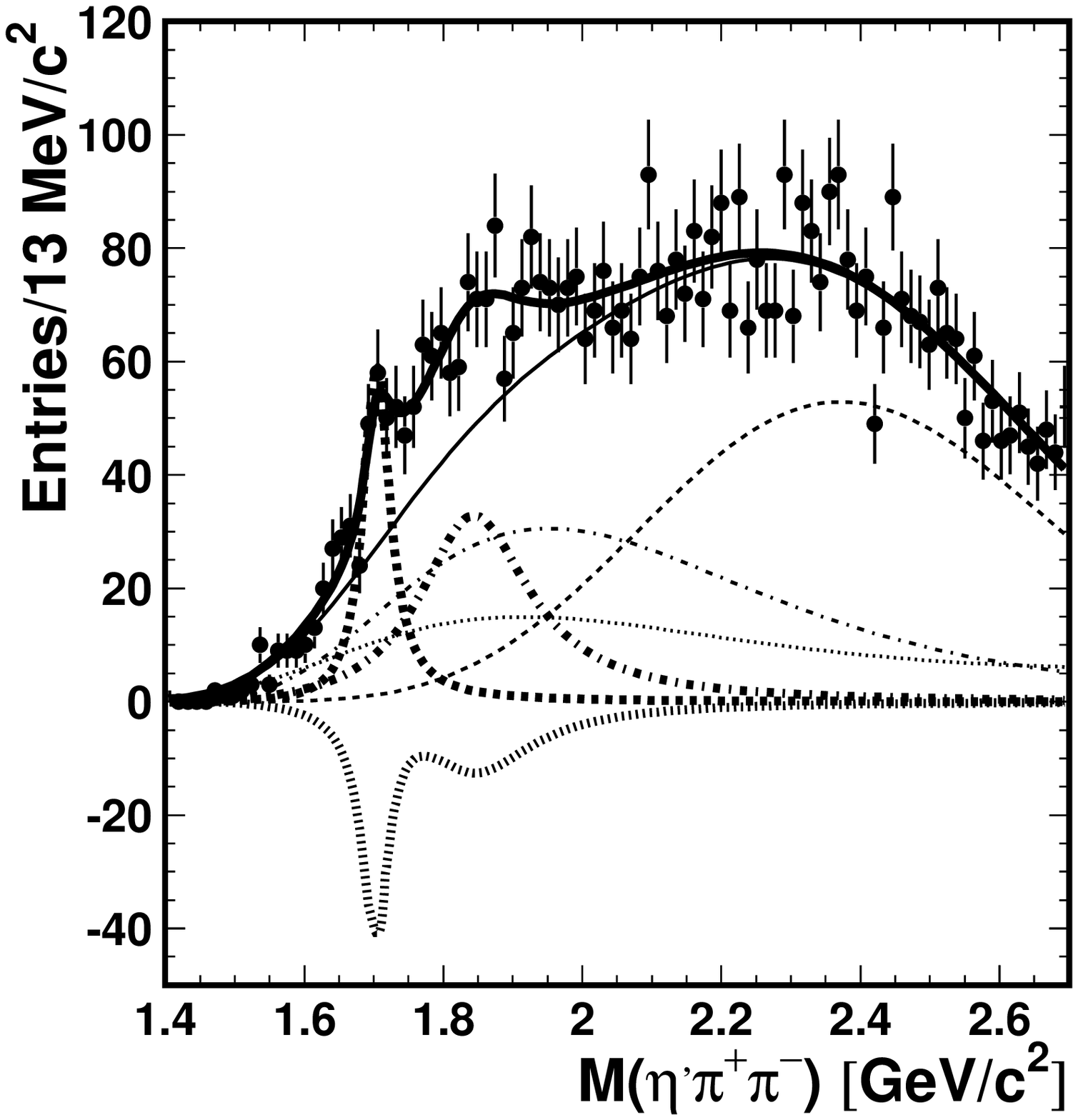}
\end{center}
\caption{Results of a combined fit for the $X(1835)$ and $\eta(1760)$
with interference between them.
The points with error bars are data.
The thick solid line is the fit; the thin solid line is the total
background.
The thick dashed (dot-dashed, dotted) line is the fitted signal for the
$\eta(1760)$ ($X(1835)$, the interference term between them).
The left (right) panel represents the solution with
constructive (destructive) interference.
}\label{Fig fit result for x1835}
\end{figure}

The resulting $\etap\pip\pim$ invariant mass distribution
was shown in Fig.~\ref{Fig fit result for x1835}. According to existing observations, two resonances,
$X(1835)$ and $\eta(1760)$, have been reported
in the lower mass region above the $\eta^\prime\pip\pim$ threshold.
A fit with the $X(1835)$ and $\eta(1760)$ signals plus their interference
is performed to the lower-mass events. Here, the $X(1835)$ mass and width are fixed at the BES value.
There are two solutions with equally good fit quality;
the results are shown in Fig.~\ref{Fig fit result for x1835}.
In either solution,
the statistical significance is
$2.9\sigma$ for  the $X(1835)$ and $4.1\sigma$ for the $\eta(1760)$.
Upper limits on the product $\Gamma_{\GG} \BR(\etap\pip\pim)$ for the $X(1835)$
at the $90\%$ C.L. are determined to be
$35.6$ eV$/c^2$ and $83$ eV$/c^2$
for the constructive- and destructive-interference solutions,
respectively.

C-even glueballs can be studied in the process
$e^+e^- \to\gamma^{*} \to H+\mathcal{G}_J$,
where $H$ denotes a $c\bar{c}$ quark pair or charmonium state
and $\mathcal{G}_J$ is a glueball.
So if the $X(1835)$ was a candidate of glueball, it can also be searched for in the process
$e^+e^-\to J/\psi X(1835)$ at $\sqrt{s}\approx10.6$~GeV at Belle
using a data sample of 672 fb$^{-1}$.

\begin{figure}
\begin{center}
\includegraphics[scale=0.14]{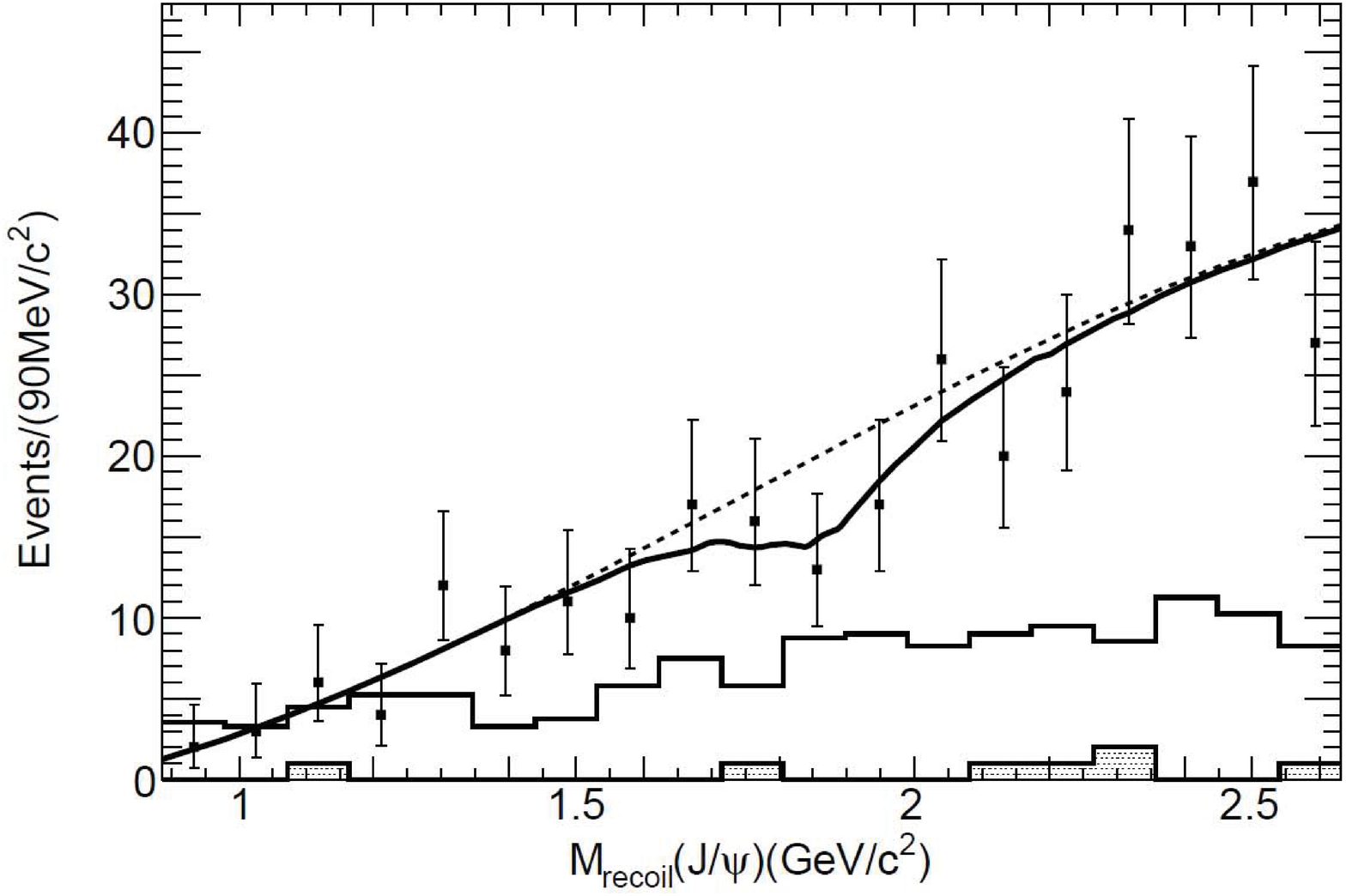}
\includegraphics[scale=0.14]{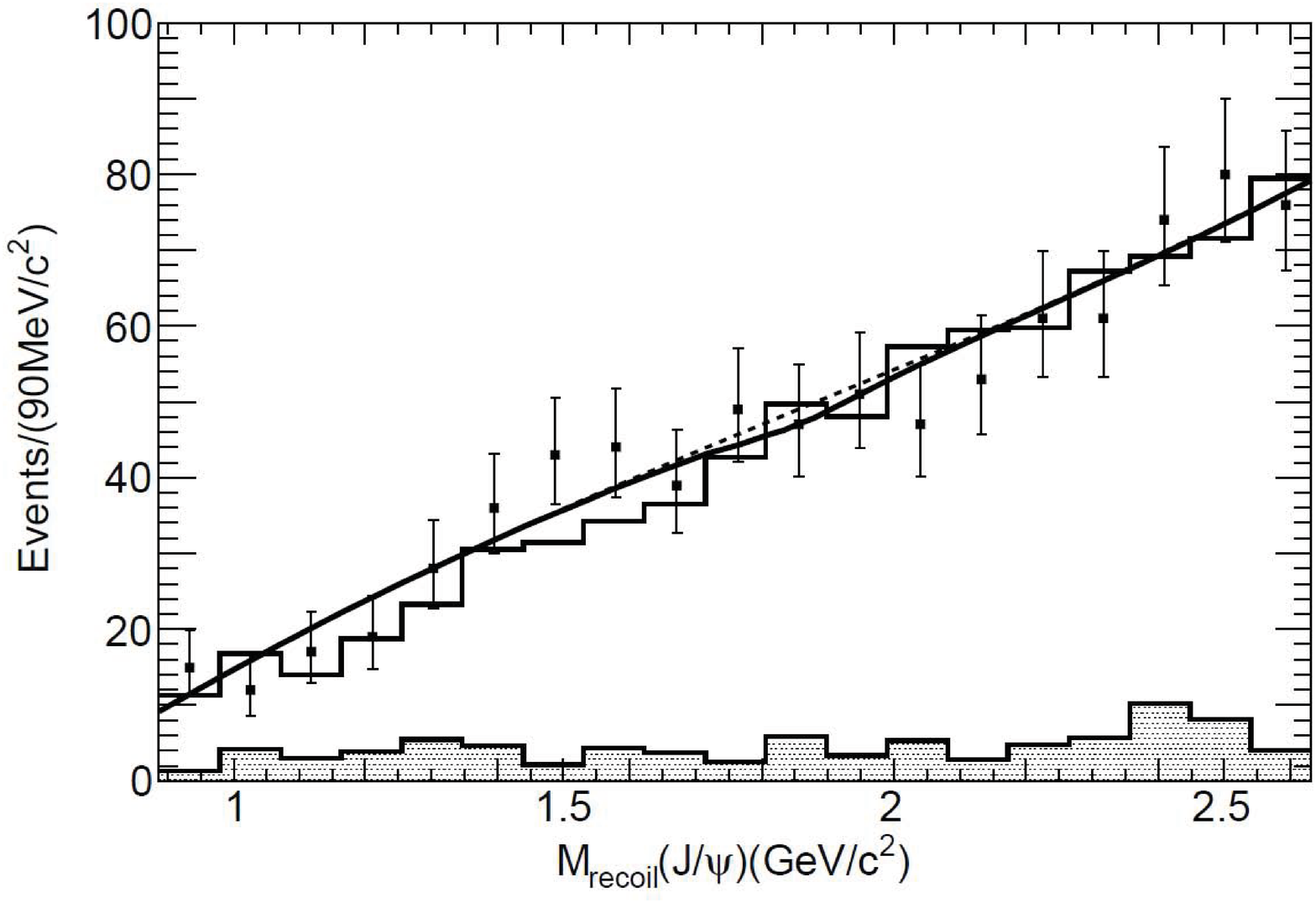}
\put(-300,90){(a)}
\put(-125,90){(b)}
\end{center}
\caption{\label{fitx} The data points are for the distributions of
the recoil mass against $J/\psi$ reconstructed from (a) $\mu^+\mu^-$ and (b) $e^+e^-$.
The histograms represent the backgrounds
from the $J/\psi$ sideband; the hatched histograms represent
charmed- plus $uds$-quark backgrounds.
The solid lines are results of the fits and the dashed lines are background shapes.}
\end{figure}

After all the event selections, the $M_{\rm recoil}$ distributions of the $\jpsi$
are shown in Fig.~\ref{fitx}. An unbinned simultaneous  maximum likelihood
fit to the $M_{\rm recoil}$ distributions was performed for the $\mu^+\mu^-$ and $e^+e^-$ channels
in the region of 0.8 GeV/c$^2<M_{\rm recoil}<2.8$ GeV/$c^2$, which constrains the expected signal from $J/\psi \to \mu^+\mu^-$
and $J/\psi \to e^+e^-$ to be consistent with the ratio of
$\varepsilon_i$ and ${\cal B}_i$,
where $\varepsilon_i$ and ${\cal B}_i$ are the efficiency and branching
fraction for the two channels, respectively.
No significant evidence of $X(1835)$ is found, and an
upper limit is set on its cross section times the branching fraction:
$\sigma_{\rm Born}(e^+e^- \to J/\psi X(1835)) \cdot$
{${\cal B}(X(1835)\to \ >2$ charged tracks)} $< 1.3 \ {\rm fb}$ at 90\%
C.L. This upper limit is three orders of magnitude smaller than the cross section of
prompt production of $J/\psi$.
According to this work, no evidence was found to support
the hypothesis of $X(1835)$ to be a glueball produced with
$J/\psi$ in the Belle experiment.

\section{$\eta\eta$ mass spectra}

According to lattice QCD predictions, the
lowest mass glueball with $J^{PC}=0^{++}$ is in the mass region from $1.5$
to $1.7$~GeV/$c^2$.  However, the mixing of the pure glueball with
nearby $q \bar q$ nonet mesons makes the identification of the
glueballs difficult in both experiment and theory.
Radiative $J/\psi$ decay is a gluon-rich process and
has long been regarded as one of the most promising hunting grounds for
glueballs. In particular, for a $J/\psi$ radiative decay to two
pseudoscalar mesons, it offers a very clean laboratory to search for
scalar and tensor glueballs because only intermediate states with
$J^{PC}=even^{++}$ are possible.

\begin{figure*}[htbp]
   \vskip -0.1cm
   \centering
   {\includegraphics[height=4.5cm]{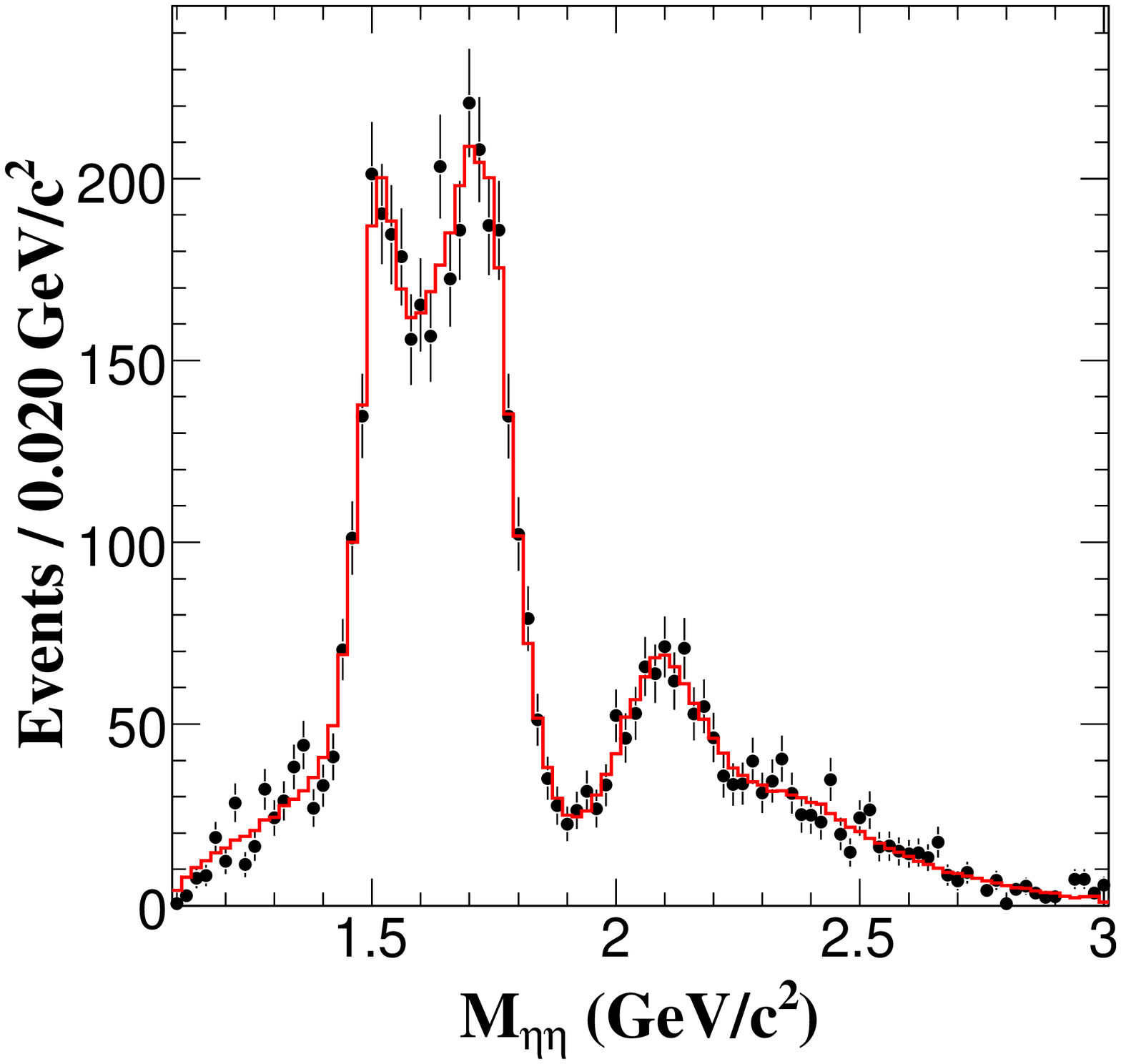}
    \put(-130,7){(a)}\put(-75,100){$\chi^{2}/N_{bin}$$=$$1.72$}}
   {\includegraphics[height=4.5cm]{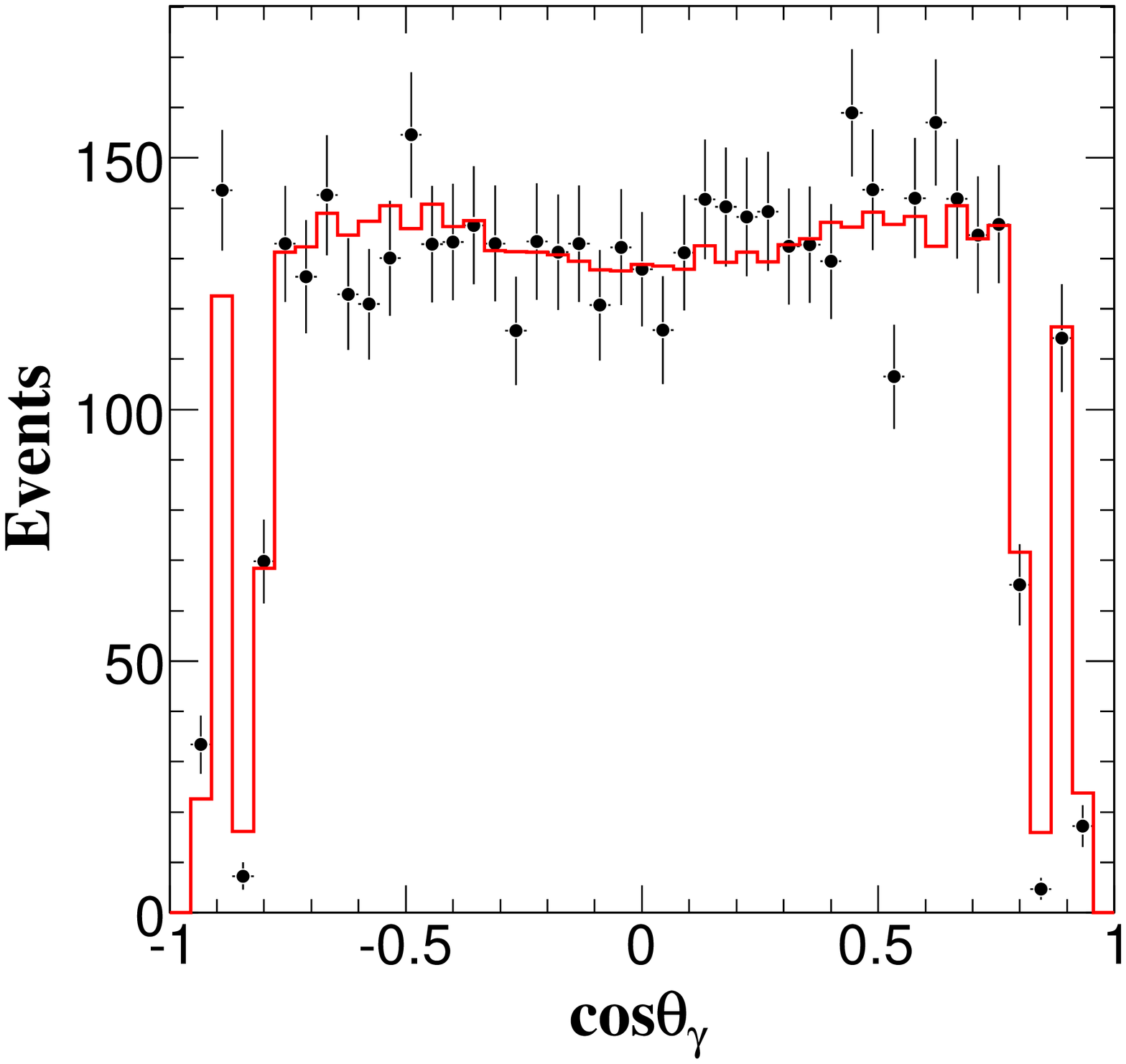}
    \put(-130,7){(b)}\put(-100,70){$\chi^{2}/N_{bin}$$=$$1.19$}}
   {\includegraphics[height=4.5cm]{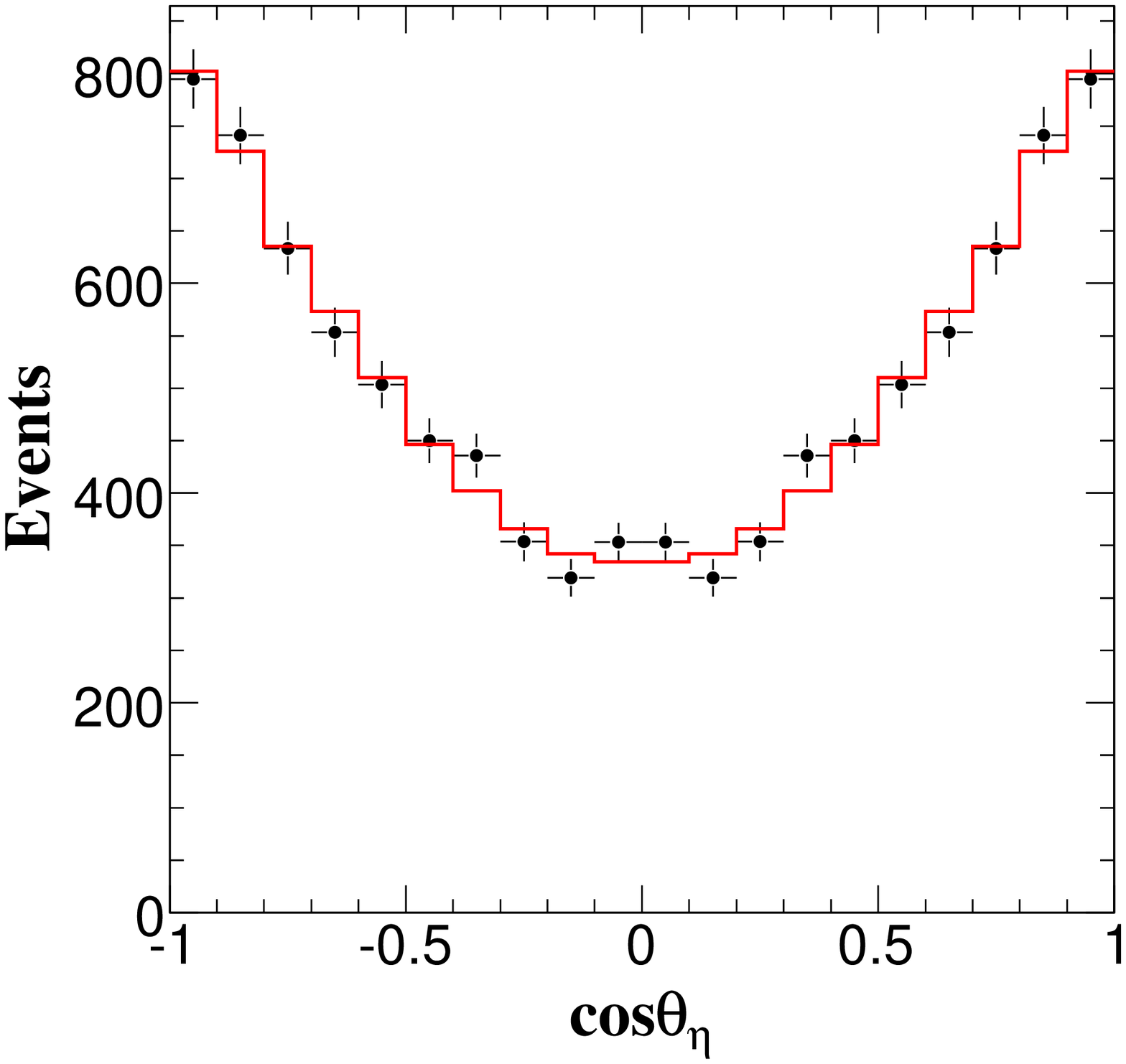}
    \put(-130,7){(c)}\put(-100,110){$\chi^{2}/N_{bin}$$=$$0.69$}}
   {\includegraphics[height=4.5cm]{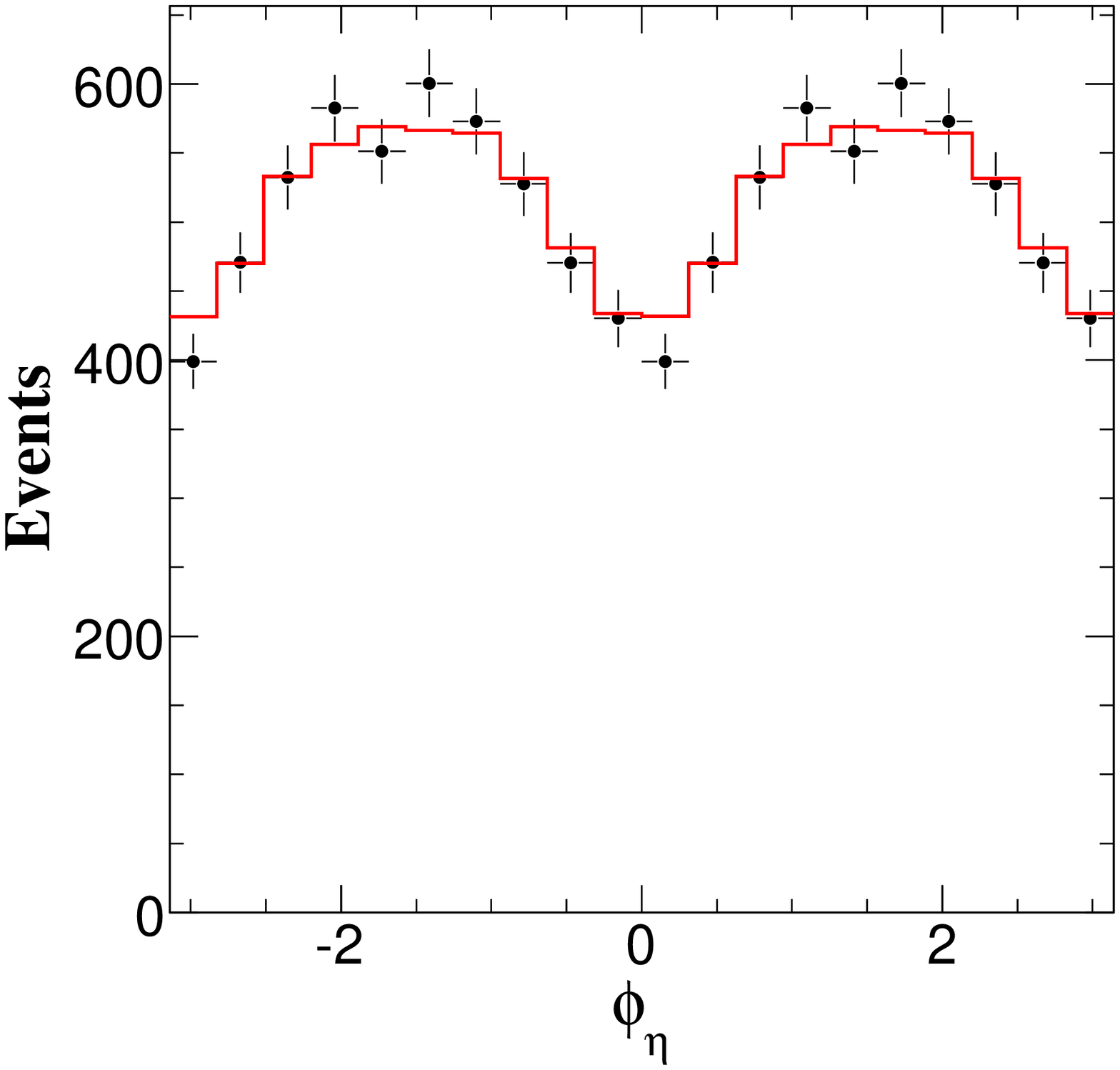}
    \put(-130,7){(d)\put(30,60){$\chi^{2}/N_{bin}$$=$$0.68$}}
   }
   \caption{Comparisons between data and PWA fit projections: (a) the
     invariant mass spectrum of $\eta\eta$, (b)-(c) the polar angle of
     the radiative photon in the $J/\psi$ rest frame and $\eta$ in the
     $\eta\eta$ helicity frame, and (d) the azimuthal angle of $\eta$ in
     the $\eta\eta$ helicity frame. The black dots with error bars are
     data with background subtracted, and the solid histograms show the
     PWA projections.}
   \vskip -0.5cm
   \label{fig:pwafitresult}
\end{figure*}

Recently the study of $J/\psi \to
\gamma \eta \eta$ was made by BESIII using
$2.25\times 10^{8}$ $J/\psi$ events~\cite{bes3-2eta},
where the $\eta$ meson is detected in its $\gamma\gamma$
decay. There are six resonances, $f_{0}(1500)$, $f_{0}(1710)$,
$f_{0}(2100)$, $f_{2}^{'}(1525)$, $f_{2}(1810)$, $f_{2}(2340)$, as
well as $0^{++}$ phase space and $J/\psi\to\phi\eta$ included in the
basic solution. The masses and widths of
the resonances, branching ratios of $J/\psi$ radiative decaying to X
and the statistical significances are summarized in Table~\ref{mwb}.
The comparisons of the $\eta\eta$ invariant mass spectrum,
$\cos\theta_{\eta}$, $\cos\theta_{\gamma}$ and $\phi_{\eta}$
distributions between the data and the partial wave analysis (PWA) fit projections
are displayed in Fig.~\ref{fig:pwafitresult}.
The results show that the
  dominant $0^{++}$ and $2^{++}$ components are from
  the $f_0(1710)$, $f_0(2100)$, $f_0(1500)$, $f_2'(1525)$, $f_2(1810)$ and $f_2(2340)$.

\begin{table}[ph]
\tbl{Summary of the PWA results, including the masses and widths for resonances, branching ratios of
$J/\psi\to\gamma$X, as well as the significance. The first errors are statistical and the second ones are systematic.
The statistic significances here are obtained according to the changes of the log likelihood.}
{\begin{tabular}{ccccc}
\hline\hline Resonance  &Mass(MeV/$c^{2}$) &Width(MeV/$c^{2}$)  &$\BR{(J/\psi\to\gamma X\to\gamma \eta\eta)}$ &Significance\\ \hline

$f_{0}(1500)$  &1468$^{+14+23}_{-15-74}$  &136$^{+41+28}_{-26-100}$    &$(1.65^{+0.26+0.51}_{-0.31-1.40})\times10^{-5}$  &8.2~$\sigma$   \\

$f_{0}(1710)$  &1759$\pm6^{+14}_{-25}$    &172$\pm10^{+32}_{-16}$      &$(2.35^{+0.13+1.24}_{-0.11-0.74})\times10^{-4}$  &25.0~$\sigma$  \\

$f_{0}(2100)$  &2081$\pm13^{+24}_{-36}$   &273$^{+27+70}_{-24-23}$     &$(1.13^{+0.09+0.64}_{-0.10-0.28})\times10^{-4}$  &13.9~$\sigma$  \\

$f_{2}^{'}(1525)$  &1513$\pm5^{+4}_{-10}$  &75$^{+12+16}_{-10-8}$      &$(3.42^{+0.43+1.37}_{-0.51-1.30})\times10^{-5}$  &11.0~$\sigma$  \\

$f_{2}(1810)$  &1822$^{+29+66}_{-24-57}$  &229$^{+52+88}_{-42-155}$    &$(5.40^{+0.60+3.42}_{-0.67-2.35})\times10^{-5}$  &6.4~$\sigma$   \\

$f_{2}(2340)$  &2362$^{+31+140}_{-30-63}$  &334$^{+62+165}_{-54-100}$  &$(5.60^{+0.62+2.37}_{-0.65-2.07})\times10^{-5}$  &7.6~$\sigma$   \\\hline \hline
\end{tabular} \label{mwb}}
\end{table}

\begin{figure}[h]
\centering
\begin{minipage}{14pc}
\includegraphics[width=12pc]{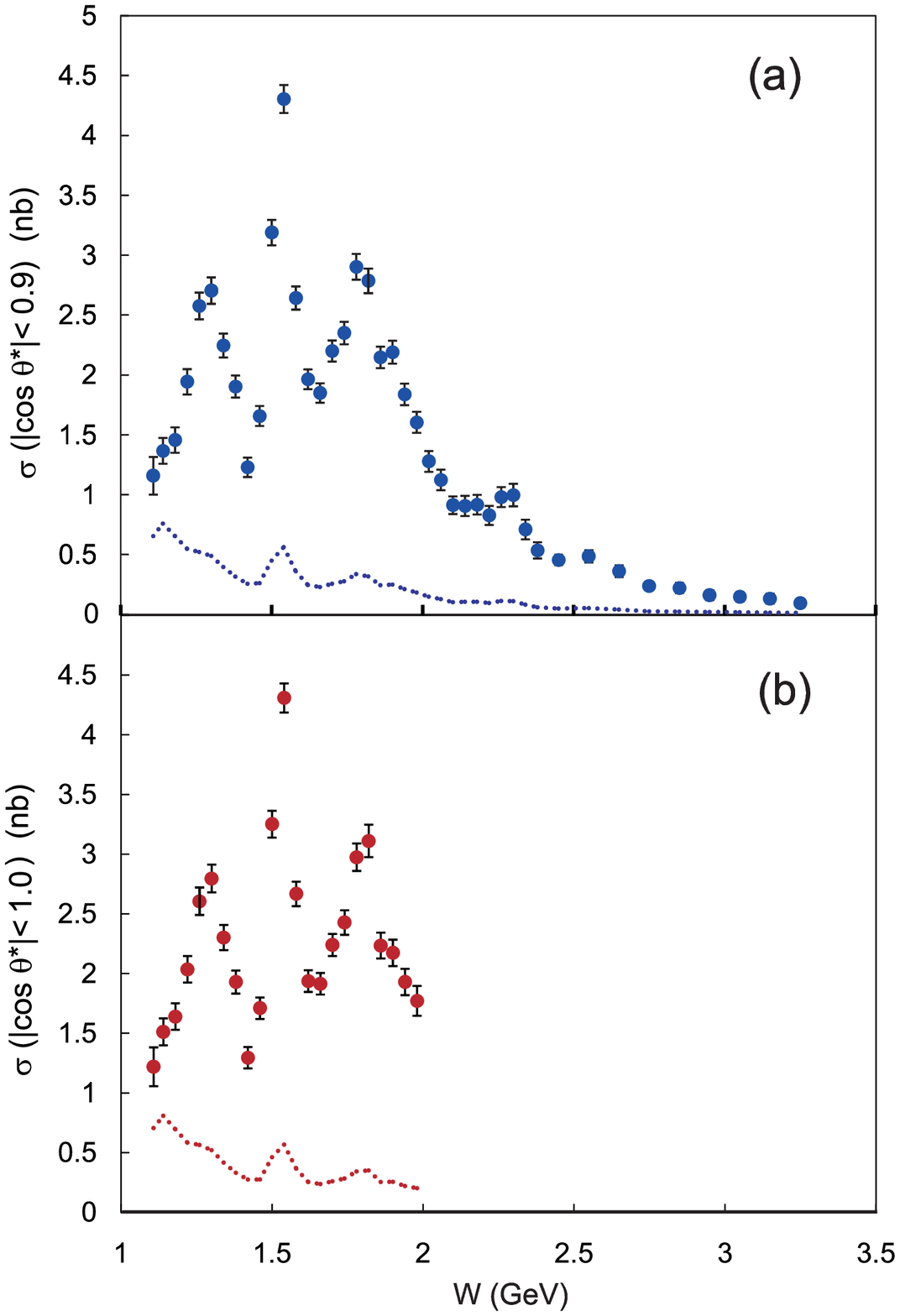}
\caption{\label{cs-gg} (a) The cross section integrated
over $|\cos \theta^{\ast}|< 0.9$ and (b) over
$|\cos \theta^\ast|< 1.0$ for $W < 2.0$~GeV. Here $\theta^\ast$
is the angle of $\eta$ in two-photon system. The dotted curve
shows the size of the systematic uncertainty.}
\end{minipage}\hspace{1.5pc}%
\begin{minipage}{14pc}
\includegraphics[width=12pc]{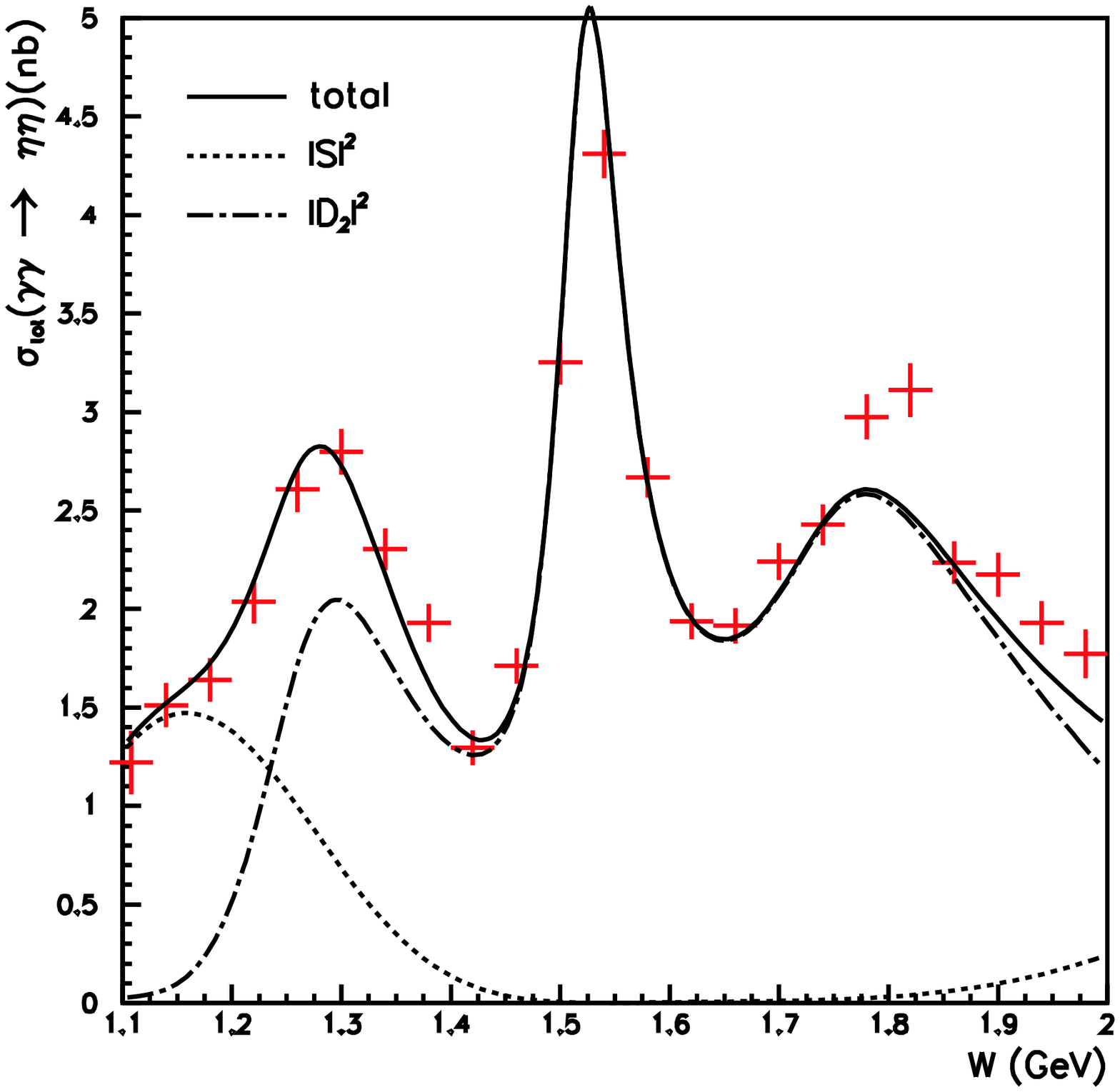}
\caption{\label{total_cs} Total cross sections and fitted curves for
 the nominal fit in the high mass region (solid curve).
 Dotted (dot-dashed) curves are $|S|^2$ ($|D_2|^2$) from the fit.}
\end{minipage}
\end{figure}

$\eta \eta$ mass spectrum was also ever studied by Belle in two-photon
process $\gamma \gamma \to \eta \eta$ using 393 fb$^{-1}$ data~\cite{belle-2eta}.
This pure neutral final states are
selected with energy sum and cluster counting triggers, both of which information are provided by
a CsI(Tl) electromagnetic calorimeter. The background was subtracted by studying sideband events in
two-dimensional $M_1(\gamma \gamma)$ versus $M_2(\gamma \gamma)$  distributions. Further background
effects are studied using $\sumpt$
distribution. Figure~\ref{cs-gg} shows the total cross sections.
For the lower energy region $1.16~\hbox{GeV} < W < 2.0$ GeV, a PWA was performed to
the differential cross section as shown in Fig.~\ref{total_cs}.
In addition to the known $f_2(1270)$ and $f_2'(1525)$,  
a tensor meson $f_2(X)$ is needed to describe $D_2$ wave, which may correspond to $f_2(1810)$ state,
and the mass, width and product of the two-photon decay width and branching fraction $\Gamma_{\gamma\gamma}B(\eta\eta)$
for $f_2(X)$ are obtained to be $1737\pm9$ MeV/$c^2$, $228^{+21}_{-20}$ MeV and $5.2^{+0.9}_{-0.8}$ eV, respectively.


\section{$\omega \omega$, $\omega \phi$ and $\phi \phi$ mass spectra}

An anomalous near-threshold enhancement,  denoted as the $X(1810)$, in the $\omega \phi$ invariant-mass spectrum
in the process $\jpsi \to \gamma \omega \phi$ was reported by the BESII experiment via PWA.
The analysis indicated that the $X(1810)$ quantum number assignment
favored $J^{PC}=0^{++}$ over $J^{PC}=0^{-+}$ or $2^{++}$ with
a significance of more than 10$\sigma$. The mass and width are
$M = 1812^{+19}_{-26}(stat.)\pm18$(syst.) MeV/$c^2$ and
$\Gamma = 105\pm20(stat.)\pm28(syst.)$ MeV/$c^2$, respectively, and the product branching
fraction ${\cal B}$($\jpsi\to\gamma$ $X(1810)$) ${\cal B}$($X(1810)$$\to\omega \phi$)
=$[2.61\pm0.27(stat.)\pm0.65(syst.)]\times10^{-4}$ was measured.

Possible interpretations for the $X(1810)$ include a tetraquark state, a hybrid,
or a glueball state etc., a dynamical effect arising from intermediate meson rescattering,
or a threshold cusp of an attracting resonance.

\begin{figure}
\begin{center}
\includegraphics[scale=0.32]{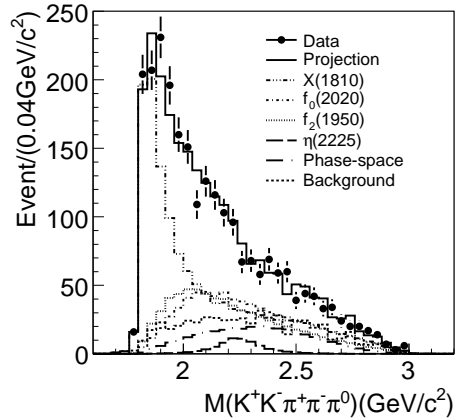}
\end{center}
\caption{\label{pwa_mwf} The $\kk\ppp$ invariant-mass distribution between data and PWA fit projections.}
\end{figure}

A PWA that uses a tensor covariant amplitude for the
$\jpsi \to \gamma \omega \phi$ process was performed again
in order to confirm  the $X(1810)$ using  $(225.3\pm2.8) \times
10^{6} \jpsi$ events~\cite{bes3-x1810}. A PWA was performed on the selected $\jpsi \to \gamma \omega \phi$
candidate events to study the properties of the $\omega \phi$ mass threshold enhancement.
In the PWA,  the enhancement is denoted as $X$, and the
decay processes are described with sequential 2-body or 3-body decays:
$\jpsi\to\gamma X, X\to\omega \phi$, $\omega\to\ppp$ and $\phi\to\kk$. The amplitudes
of the 2-body or 3-body decays are constructed with a covariant tensor
amplitude method. Finally, together with the contributions of the $X(1810)$ and phase-space, additional
needed components are listed in Table~\ref{optimalres} for the best solution of the PWA fit.
The $J^{PC}=0^{++}$ assignment for the $X(1810)$ has by far
the highest log likelihood value among the different $J^{PC}$ hypotheses,
and the statistical significance of the $X(1810)$ is more than 30$\sigma$.
The mass and width of the $X(1810)$
are determined to be $M=1795\pm7(stat.)^{+13}_{-5}(syst.)\pm19(mod.)$ MeV/$c^2$ and
$\Gamma=95\pm10(stat.)^{+21}_{-34}(syst.)\pm75(mod.)$ MeV/$c^2$ and the product branching fraction is measured to be
${\cal B}(\jpsi\to\gamma X(1810))\times{\cal B}(X(1810)\to\omega \phi)=(2.00\pm0.08(stat.)^{+0.45}_{-1.00}(syst.)\pm1.30(mod.))\times10^{-4}$.
The contributions of each component of the best solution of the PWA fit
are shown in Fig.~\ref{pwa_mwf}. The enhancement is not compatible with being due either
to the $X(1835)$ or the $X(p\bar{p})$, due to the different mass and spin-parity.
The search for other possible states
decaying to $\omega \phi$ would be interesting.

\begin{table}[ph]
\tbl{Results from the best PWA fit solution.}
{\begin{tabular}{cccccc}
\hline\hline
Resonance&J$^{PC}$&M(MeV$/c^2$)&$\Gamma$(MeV$/c^2$)&Events&Significance\\\hline
$X(1810)$&0$^{++}$&$1795\pm7$&$95\pm10$&$1319\pm52$&$>30\sigma$\\\hline
f$_{2}$(1950)&2$^{++}$&1944&472&$665\pm40$&20.4$\sigma$\\\hline
f$_{0}$(2020)&0$^{++}$&1992&442&$715\pm45$&13.9$\sigma$\\\hline
$\eta(2225)$&0$^{-+}$&2226&185&$70\pm30$&$6.4\sigma$\\\hline
phase space&0$^{-+}$&--- &--- &$319\pm24$&9.1$\sigma$\\\hline\hline
\end{tabular} \label{optimalres}}
\end{table}

In the two-photon processes
$\gamma\gamma\to \omega\jpsi$ and $\phi
\jpsi$, a state $X(3915)$ and an evidence for
$X(4350)$ were observed.
It is very natural to extend the above theoretical picture to similar
states coupling to $\op$, since the only difference between such
states and the $X(3915)$ or $X(4350)$ is
the replacement of the $c\bar{c}$ pair with a pair of light quarks.
States coupling to $\oo$ or $\phi\phi$ could also provide information on the
classification of the low-lying states coupled to pairs of light vector
mesons.

\begin{figure}
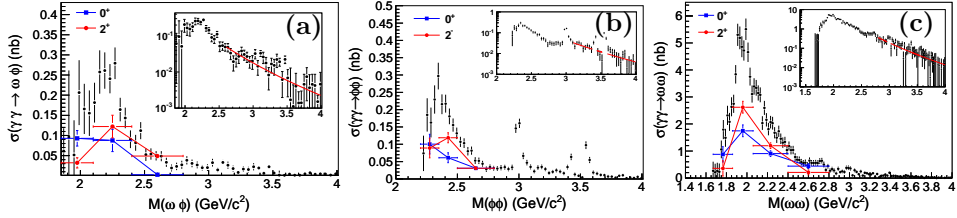

\begin{center}
\includegraphics[width=1.1in, angle=-90]{fig2a.epsi}
\includegraphics[width=1.1in, angle=-90]{fig2b.epsi}
\includegraphics[width=1.1in, angle=-90]{fig2c.epsi}
 \put(-257,-12){ \bf (a)}
 \put(-140,-11){ \bf (b)}
 \put(-25,-11){ \bf (c)}
 \end{center}
\caption{The cross sections of $\gamma \gamma \to \omega \phi$
(a), $\phi \phi$ (b), and $\omega \omega$ (c)
are shown as points with error
bars. The fraction contributions for different $J^P$ values as a
function of $M(VV)$  are shown as the points and squares with error bars.} \label{cross-section}
\end{figure}

The $\gamma \gamma \to VV$ cross sections are shown in
Fig.~\ref{cross-section}~\cite{gg2vv}.
The fraction of cross sections for
different $J^P$ values as a function of $M(VV)$ is also shown in
Fig.~\ref{cross-section}. We conclude that there are at least two
different $J^P$ components ($J=0$ and $J=2$) in each of the three
final states. The inset also shows the distribution of the cross
section on a semi-logarithmic scale, where, in the high
energy region, we fit the $W^{-n}_{\gamma \gamma}$ dependence of
the cross section.

We observe
clear structures at $M(\op)\sim 2.2$~GeV/$c^2$, $M(\phi \phi)\sim 2.35$~GeV/$c^2$,
and $M(\oo)\sim 2.0$~GeV/$c^2$. While there are substantial spin-zero components in
all three modes, there are also spin-two components near threshold.

\section{Conclusion}

I have reviewed some results on the charmonium and light hadron spectroscopy
mainly from BESIII and Belle experiments,
including the observation of $\psi(4040)/\psi(4160) \to \eta \jpsi$,
some measurements on the $\eta_c/\eta_c(2S)$ resonance parameters and their decays,
the evidence of the $\psi_2(1^3D_2)$ state in the $\chi_{c1}\gamma$ mass spectrum,
the X(1835) research in more processes, and the analysis of the $\eta \eta$,
$\omega \phi$, $\phi\phi$ and $\omega \omega$ mass spectra.

\section*{Acknowledgments}

This work is supported partly by the Fundamental Research Funds for the Central Universities of China (303236).


\end{document}